\documentclass[runningheads]{llncs}
\usepackage{graphicx}

\usepackage{url}
\usepackage[colorlinks,bookmarks=false]{hyperref}
\usepackage[usenames,dvipsnames, table]{xcolor}
\hypersetup{citecolor=blue,linkcolor=blue}

\usepackage[numbers]{natbib}

\usepackage{wrapfig, lipsum,booktabs}
\usepackage{xspace}
\usepackage{listings}
\usepackage{booktabs}
\usepackage{enumitem}
\usepackage[final]{pdfpages}
\usepackage{multicol}
\usepackage{multirow}
\usepackage{lipsum}
\usepackage{mwe}
\usepackage{color,colortbl}
\usepackage{mdframed}
\usepackage{blindtext}
\usepackage{subcaption}
\usepackage[utf8]{inputenc}
\usepackage[export]{adjustbox}
\usepackage{wrapfig}
\usepackage{tabu}
\usepackage{makecell}
\usepackage{footmisc}
\usepackage{changepage}
\usepackage{array}
\usepackage{amsfonts}
\usepackage{amsmath}
\usepackage{float}
\usepackage[linesnumbered,algoruled,boxed,lined]{algorithm2e}

\usepackage{longtable}
\usepackage{lipsum} 

\SetKw{Continue}{continue}
\SetKw{Break}{break}
\SetKw{Break}{}
\SetKwInOut{Input}{Input}
\SetKwInOut{Optional}{Optional}
\SetKwInOut{Output}{Output}
\SetAlFnt{\small}
\SetAlCapFnt{\small}
\SetAlCapNameFnt{\small}

\SetCommentSty{mycommfont}

\newcommand{\jun}[1]{{{\textcolor{red}{\textbf{Jun:}}}{\textcolor{red}{\textbf{#1}}}}}

\newcommand{\yuchen}[1]{{{\textcolor{red}{\textbf{Yuchen:}}}{\textcolor{cyan}{\textbf{#1}}}}}

\newcommand{\afl}{{\sc AFL}\xspace}
\newcommand{\aflcmin}{{\sc AFL-CMIN}\xspace}
\newcommand{\qsym}{{\sc QSYM}\xspace}
\newcommand{\pafl}{{\sc PAFL}\xspace}
\newcommand{\pfuzz}{{\sc P-FUZZ}\xspace}

\newcommand{\afledge}{{\sc AFL-EDGE}\xspace}

\newenvironment{ditemize}{%
\begin{list}{{\bf $\bullet$}}{%
\setlength{\labelindent}{\parindent}%
\setlength{\itemsep}{0.3em}%
\setlength{\leftmargin}{2em}}}{
\end{list}}

\newcommand{\ie}{i.e.,\xspace}
\newcommand{\eg}{e.g.,\xspace}
\newcommand{\etal}{et al.\xspace}

\newcommand{\vs}{v.s.\xspace}

\newcommand{\bugnum}{14\xspace}

\begin{document}
\title{Facilitating Parallel Fuzzing with Mutually-exclusive Task Distribution} 
\author{}
\institute{}
%
%
\author{Yifan Wang\inst{1}\thanks{These authors contributed equally.} \and
Yuchen Zhang \inst{1}\protect\footnotemark[1] \and
Chenbin Pang\inst{2}\thanks{This work was done while Pang was a Visiting Scholar at Stevens Institute of Technology.} \and 
Peng Li\inst{3} \and  
Nikolaos Triandopoulos\inst{1} \and
Jun Xu\inst{1}
}
\authorrunning{F. Author et al.}
%
\institute{Stevens Institute of Technology\and
Nanjing University
\and
ByteDance\\
}
\maketitle              



\begin{abstract} Fuzz testing, or fuzzing, has become one of the de facto standard techniques for bug finding in the software industry. In general, fuzzing provides various inputs to the target program with the goal of discovering un-handled exceptions and crashes. In business sectors where the time budget is limited, software vendors often launch many fuzzing instances in parallel as a common means of increasing code coverage. However, most of the popular fuzzing tools --- in their parallel mode --- naively run multiple instances concurrently, without elaborate distribution of workload. This can lead different instances to explore overlapped code regions, eventually reducing the benefits of concurrency. In this paper, we propose a general model to describe parallel fuzzing. This model distributes mutually-exclusive but similarly-weighted tasks to different instances, facilitating concurrency and also fairness across instances. Following this model, we develop a solution, called \afledge, to improve the parallel mode of \afl, considering \emph{a round of mutations to a unique seed} as a task and adopting edge coverage to define the uniqueness of a seed. We have implemented \afledge on top of \afl and evaluated the implementation with \afl on 9 widely used benchmark programs. It shows that \afledge can benefit the edge coverage of \afl. In a 24-hour test, the increase of edge coverage brought by \afledge to \afl ranges from 9.5\% to 10.2\%, depending on the number of instances. As a side benefit, we discovered \bugnum previously unknown bugs.



\end{abstract}
\section{Introduction}\label{sec:introduction}

Thanks to its direct and easy application to production-grade software without human aids, fuzzing is gaining tremendous popularity for security testing. In today's business sectors,  software systems are having shorter testing cycles~\cite{Top10Sof73:online}, and therefore, the efficiency of code coverage becomes a critically desired property of fuzzing. 

To escalate code coverage efficiency, there are two orthogonal strategies, one improving algorithms of fuzzing tools and one launching many fuzzing instances in parallel. The research community has intensively investigated the first strategy. Efforts along this line have revolutionized fuzzing from being program-structure-agnostic and black-box~\cite{aitel2002introduction,beizer1995black,myers2011art} to be program-structure-aware and grey-box/white-box~\cite{afltech,serebryany2015libfuzzer,sage,qsyminsu,stephens2016driller}, which significantly improved fuzzing efficiency by overcoming common barriers of code coverage.

However, the second strategy has been less studied and insufficiently developed. Existing fuzzing tools (\emph{e.g.},~\cite{serebryany2015libfuzzer,angora}) primarily follow American Fuzzy Lop (AFL)~\cite{afltech} to implement their parallel mode. Technically speaking, they run multiple identical instances in parallel. Depending on the implementation, different instances may either share the same group of seed inputs (or \emph{seeds})~\cite{angora,serebryany2015libfuzzer} or use separate groups of seed inputs but make periodical exchanges~\cite{afltech}. This type of parallel fuzzing, due to lack of synchronizations, leads different instances to run overlapped tasks, impeding the effectiveness of concurrency. 

In this paper, we focus on unveiling the limitations of the parallel mode in existing fuzzing tools and presenting new solutions to overcome those limitations. We start with an empirical study of the parallel mode in \afl. By tracing the exploration of all instances across the fuzzing process, we discover that different instances are indeed running overlapped tasks despite many tasks remain unaccomplished (\S~\ref{subsec:motivation}). 
We further demonstrate that this type of task overlapping can lead to reduced efficiency of code coverage.

Motivated and inspired by our empirical study, we propose a general model to describe parallel fuzzing. At the high level, the model enforces three desired properties. First, it distributes mutually-exclusive tasks to different instances, preventing the occurrence of overlaps. Second, it ensures every single task to be covered by at least one instance. This avoids the loss of fuzzing tasks and the code covered by those tasks. Finally, it assigns to each instance tasks with a similar amount of workload. Otherwise, some instances will be overloaded while the other instances are under-loaded, which can eventually degrade concurrency. 

Guided by the model above, we develop a solution, called \afledge, towards facilitating the parallel mode in \afl. Our solution defines \emph{a task} is to \emph{run a round of mutations to a unique seed} and considers the control-flow edges (or \emph{edges}) covered by a seed to determine its uniqueness. During the course of fuzzing, \afledge periodically distributes seeds that carry non-overlapped and similarly-weighted tasks to different instances, meeting the properties of our model. \afledge also enforces that all the unique seeds will cover the same set of edges as the original seeds. We envision that, in this way, \afledge can properly preserve the fuzzing capacity of \afl.

We have implemented \afledge on top of \afl, and we have evaluated \afledge with \afl using 9 widely adopted benchmark programs. Our evaluation shows that \afledge can significantly reduce the overlaps and hence, benefit the code coverage. Depending on the number of instances we launch, we can averagely reduce 57.1\% - 60.3\% of the overlaps and bring a 9.5\% - 10.2\% increase in code coverage with \afl. 
Our evaluation also demonstrates that, compared to the state-of-the-art solutions of improving parallel fuzzing~\cite{liang2018pafl,song2019p}, our solution not only brings higher improvement to efficiency of edge coverage but also better preserves the capacity of the fuzzing tools. As a side benefit, \afledge triggers over 6K unique crashes, corresponding to \bugnum new bugs. 

Our main contributions are as follows.
\vspace{-1em}
\begin{itemize}
    \item We present a general model to describe parallel fuzzing. 
    \item We develop a solution to improve the parallel mode in \afl, following the guidance of our model.
    \item We have implemented our solution on top of \afl, which can seamlessly run with other fuzzing tools that also use \afl. Source code of our implementation will be made publicly available upon publication.
    \item We evaluated our solution with \afl on 9 widely used benchmark programs. It shows that our solution can effectively reduce the overlaps and increase the code coverage of \afl.
\end{itemize}
\begin{wrapfigure}{r}{0.5\textwidth}
\vspace{-2.5em}
\begin{center}
    \includegraphics[width=1\linewidth]{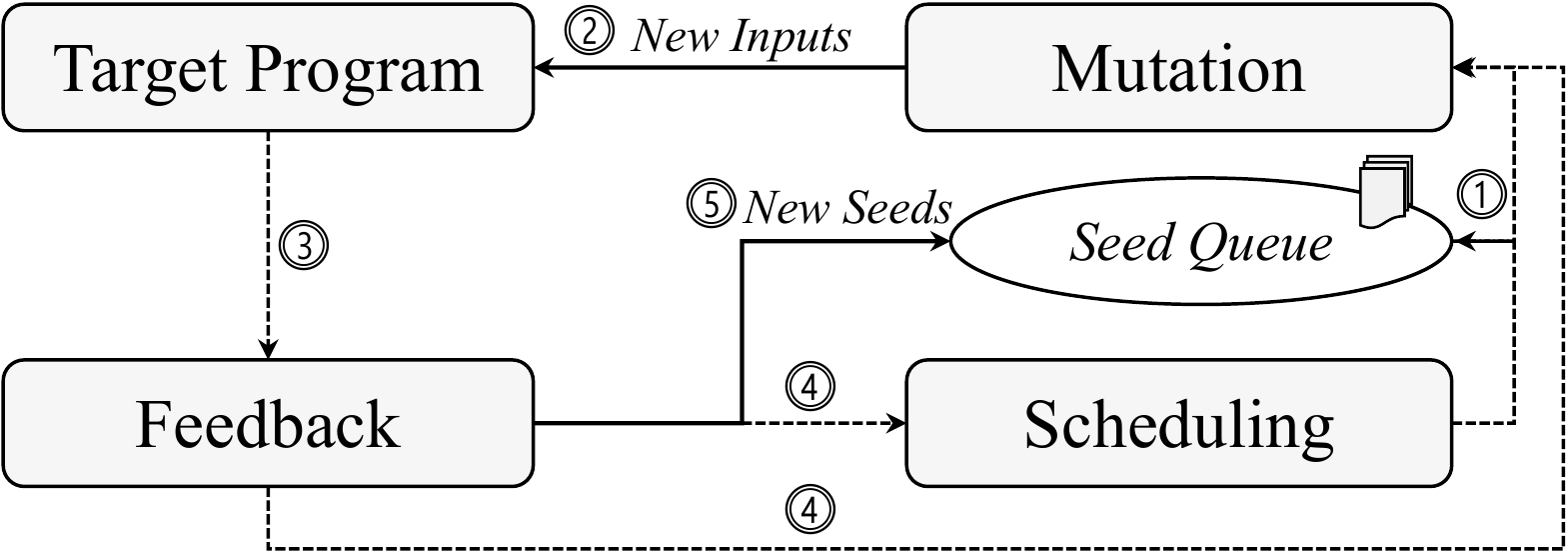}
    \caption{A general model of grey-box fuzzing.}
    \label{fig:fsm}
    \vspace{-1em}
\end{center}
\vspace{-1.5em}
\end{wrapfigure}

\section{Background and Motivation}
\label{sec:overview}

\subsection{Grey-box Fuzzing and Parallel Mode}
\label{susec:bcg}
In this research, we target grey-box fuzzing~\cite{manes2019art}, the most popular category of fuzzing. Grey-box fuzzing generally follows the \emph{feedback-scheduling-mutation} model presented in Fig.~\ref{fig:fsm}. This FSM model represents an iterative process, starting with a queue of seed inputs, or \emph{seeds}, that are typically generated from certain known test cases. In a round of fuzzing, the \emph{scheduling} picks a seed and feeds it to the \emph{mutation} process for deriving new inputs to test the target program, expecting to trigger un-handled crashes or exceptions. Both the scheduling and mutation processes are based on \emph{feedback} (\eg crashes and code coverage) obtained from the program executions on the previously generated inputs. The fuzzer also collects feedback to decide whether an input under test should be added to the seed queue.

To improve the efficiency of code coverage, many grey-box fuzzing tools~\cite{afltech,serebryany2015libfuzzer,angora} provide a parallel mode to run multiple instances concurrently. Their parallel mode mostly follows \afl. They start identically-configured instances and run them in parallel. Depending on the implementation, different instances may either share the same seed queue~\cite{serebryany2015libfuzzer} or carry separate seed queues but periodically exchange seeds~\cite{afltech}. In the latter case, each instance borrows from other instances all the seeds that bring new code coverage. While intuition suggests that more fine-grained synchronizations can benefit the effectiveness of the above parallel fuzzing, none of the existing tools carry such synchronizations.




\begin{figure}[t!]
    \small
    \renewcommand{\arraystretch}{0.5}
    \begin{tabular}{cccc}
        \includegraphics[width=0.24\linewidth]{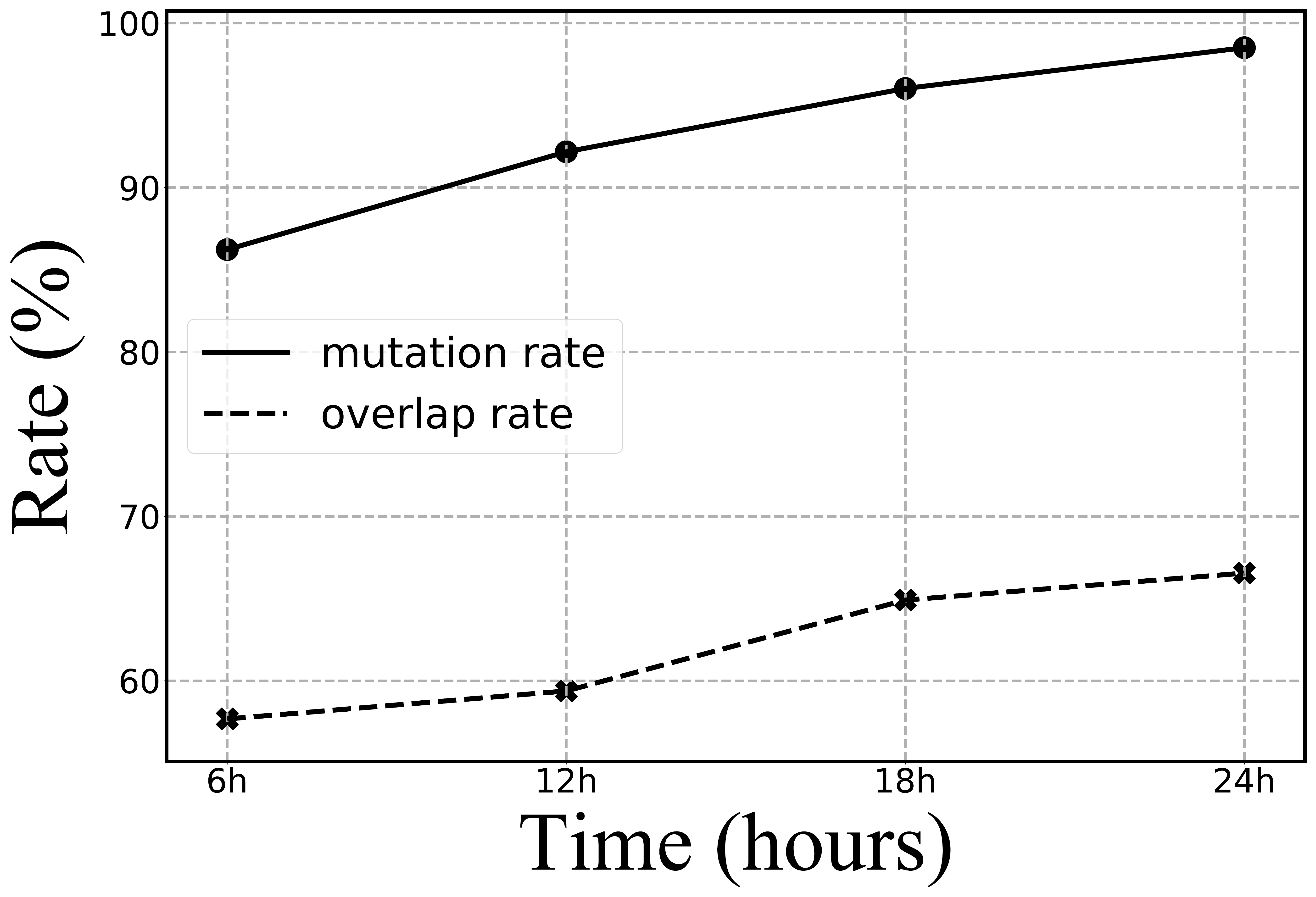} &
        \includegraphics[width=0.24\linewidth]{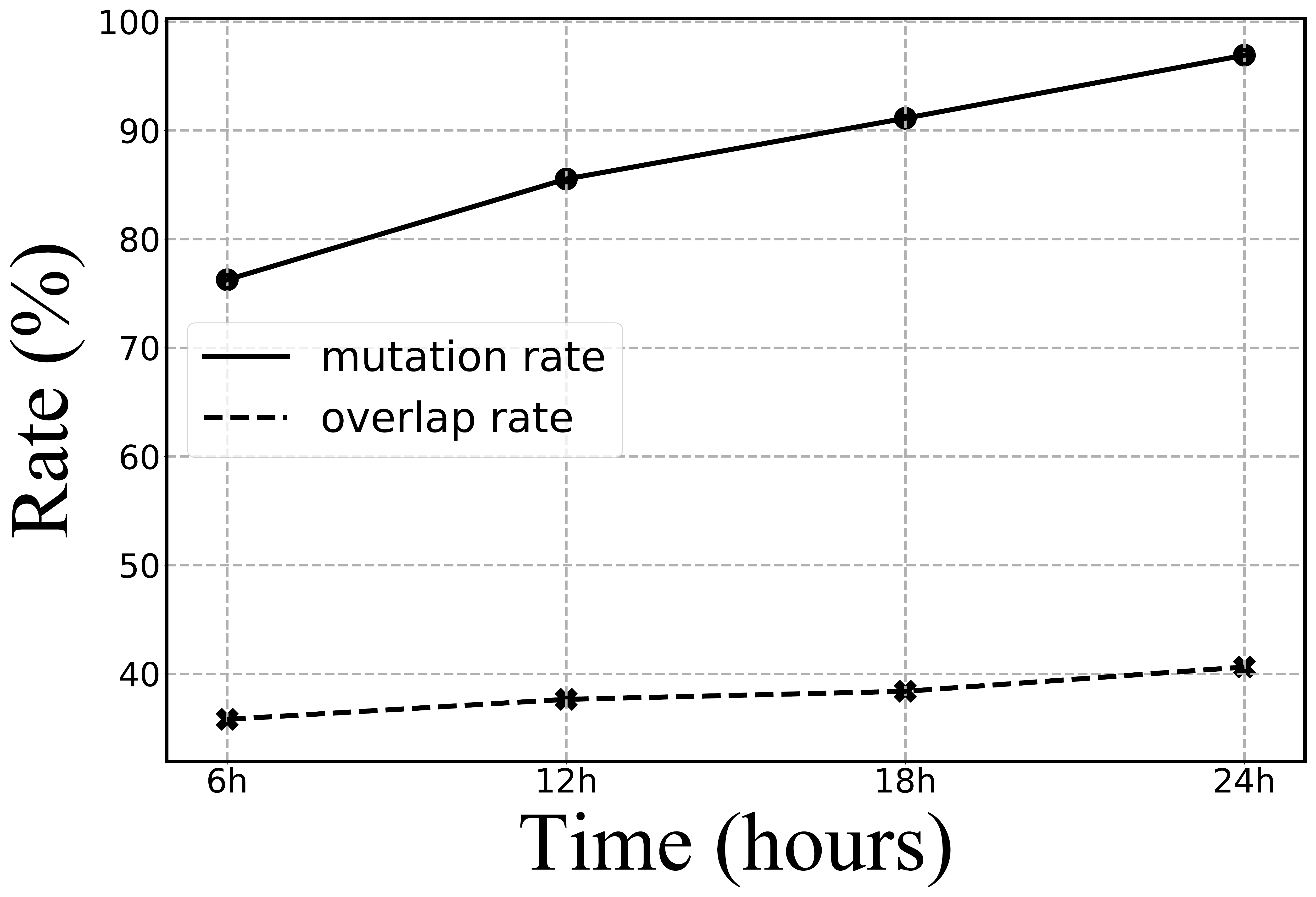} &
        \includegraphics[width=0.24\linewidth]{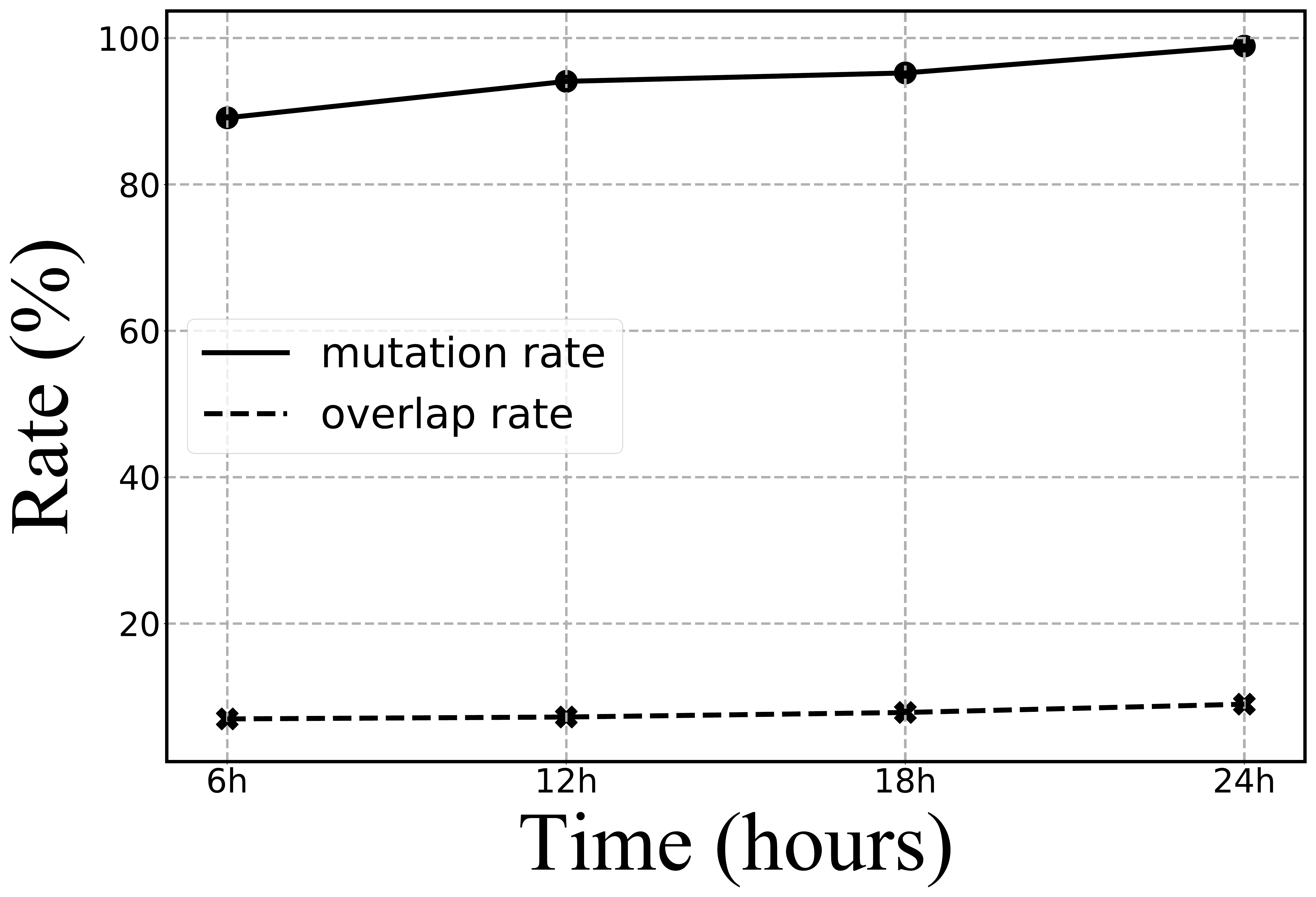} & \includegraphics[width=0.24\linewidth]{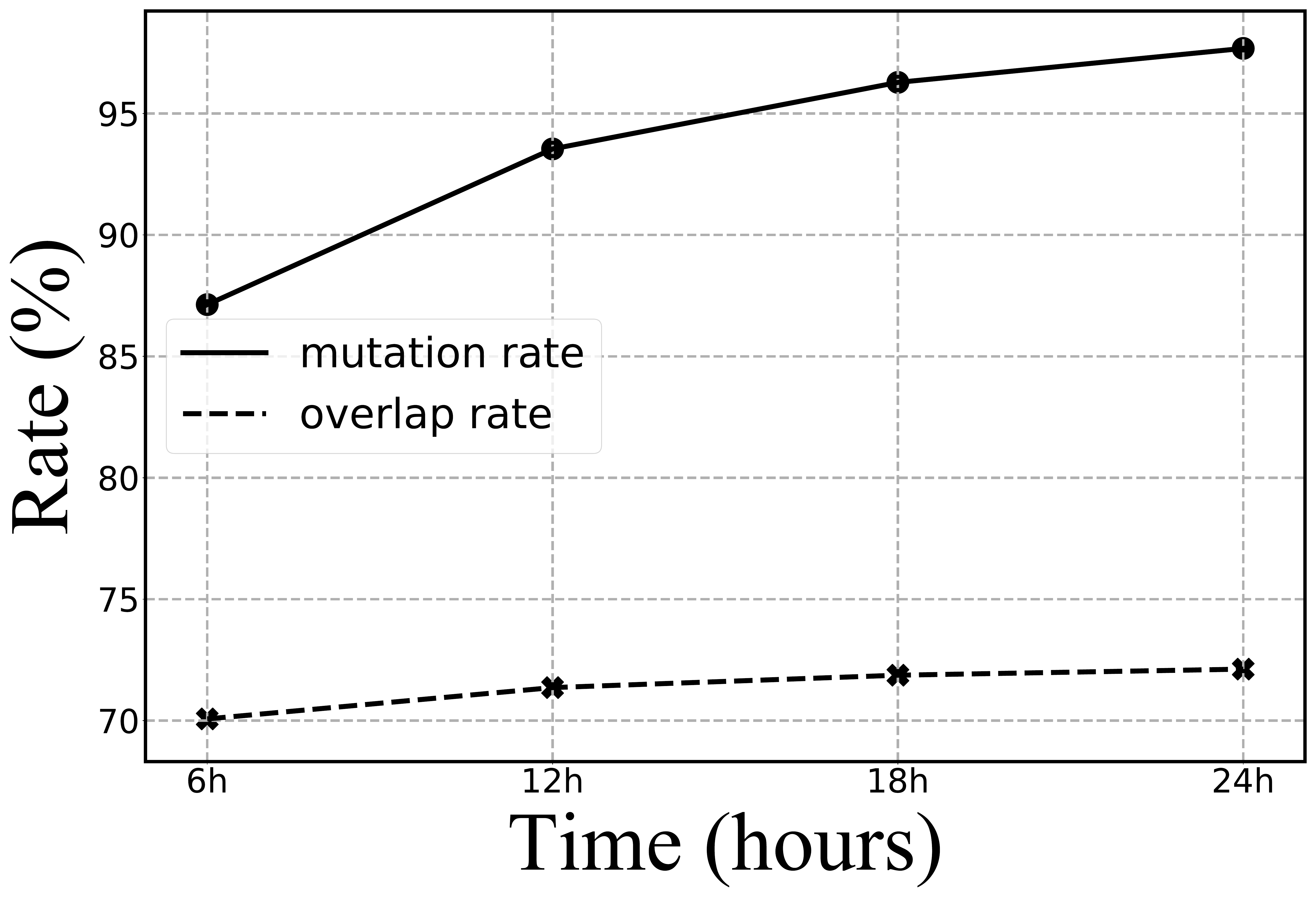} \\
         \centering{\hspace{0.6cm}(a) \sc objdump} & {\hspace{0.7cm}(b) \sc readelf} &
        \centering{\hspace{0.6cm}(c) \sc libxml} & {\hspace{0.6cm}(d) \sc nm-new} \\
    \end{tabular}
\caption{Results of measuring overlapped mutations: ``mutation rate'' - the portion of seeds mutated by at least one instance; ``overlap rate'' - the portion of seeds mutated by more than one instance.}
\label{fig:study:overlap}
\vspace{-1em}
\end{figure}

\subsection{Motivating Study \label{subsec:motivation}}
\label{susec:motivation}

Sharing seeds across instances is an intentional design of \afl~\cite{aflparal12:online}. The goal is that ``hard-to-hit but interesting test cases'' can be used by all instances to guide their work. However, intuition suggests that such a design can lead different instances to mutate the same seeds, which may eventually reduce the effectiveness of concurrency. To validate this intuition and thus, motivate our research, we perform an empirical study based on \afl. 

In our study, we run \afl on four popular benchmark programs ({\sc Objdump}, {\sc Readelf}, {\sc Libxml}, {\sc Nm-new}) with 2 parallel instances for 24 hours. We trace the mutation process to understand which seeds are mutated by which instances. We repeat the tests five times and report the average results in Fig.~\ref{fig:study:overlap}. As shown by the results, different instances are indeed mutating overlapped seeds despite many seeds are never mutated, in particular when the fuzzing time is limited. Consider the results after 6 hours as an example. On average, nearly 20\% of the seeds are never mutated. However, over 42\% of the seeds receive multiple rounds of mutations. When we increase the fuzzing time, we observe even higher rates of overlaps but still a group of non-mutated seeds. 

Although we observe overlaps among mutated seeds and the overlaps indeed delay the mutations to all seeds, they may not necessarily impede the efficiency of code coverage. This is because \afl's mutation involves random operations. In this regard, running multiple rounds of mutations to the same seed --- especially when the seed has higher potential --- may produce code coverage comparable to applying the mutations to different seeds. To verify this possibility, we run another experiment. Specifically, we randomly collect 1,000 seeds produced in the above test of each program and equally split the seeds into group \texttt{A} and group \texttt{B}. First, we run \afl to mutate group \texttt{A} for multiple rounds and we calculate the increase of code coverage after each round of mutations, where we use control flow edges as the metric of code coverage and we consider the edges covered by the original 1,000 seeds as the baseline. Then, we repeat the experiment but replace the same $X$\% of group \texttt{A} with un-mutated seeds from group \texttt{B} in each round. For example, when we set $X$ to 10, we run the original group \texttt{A} in the first round but in every following round, we replace a fixed subset of 50 seeds with other non-muted ones from group \texttt{B}. Such an experiment enables us to simulate fuzzing scenarios where a different extent of overlapped mutations happen. 

\begin{figure}[t!]
    \renewcommand{\arraystretch}{0.5}
    \begin{tabular}{cccc}
        \includegraphics[width=0.24\linewidth]{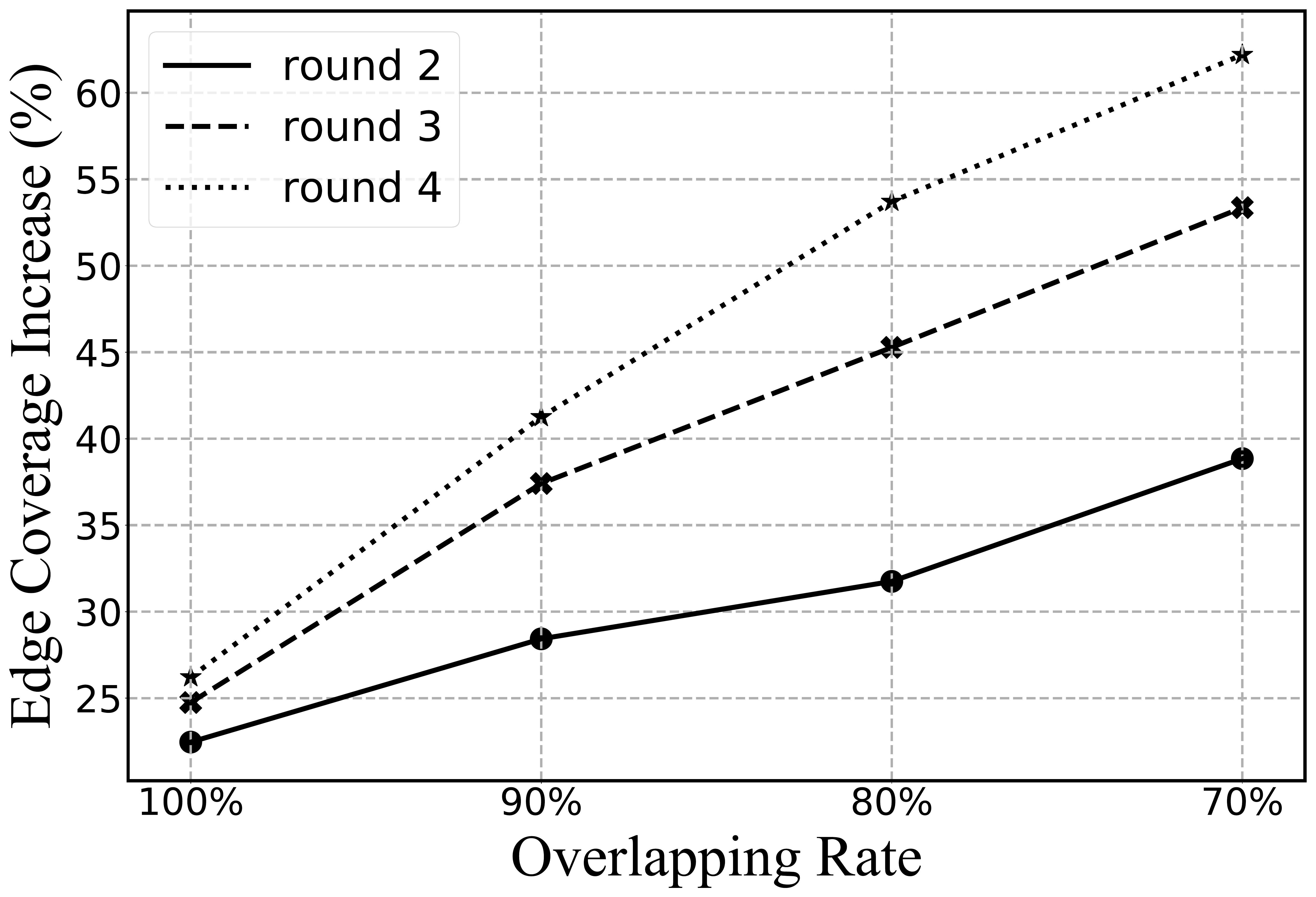} &
        \includegraphics[width=0.24\linewidth]{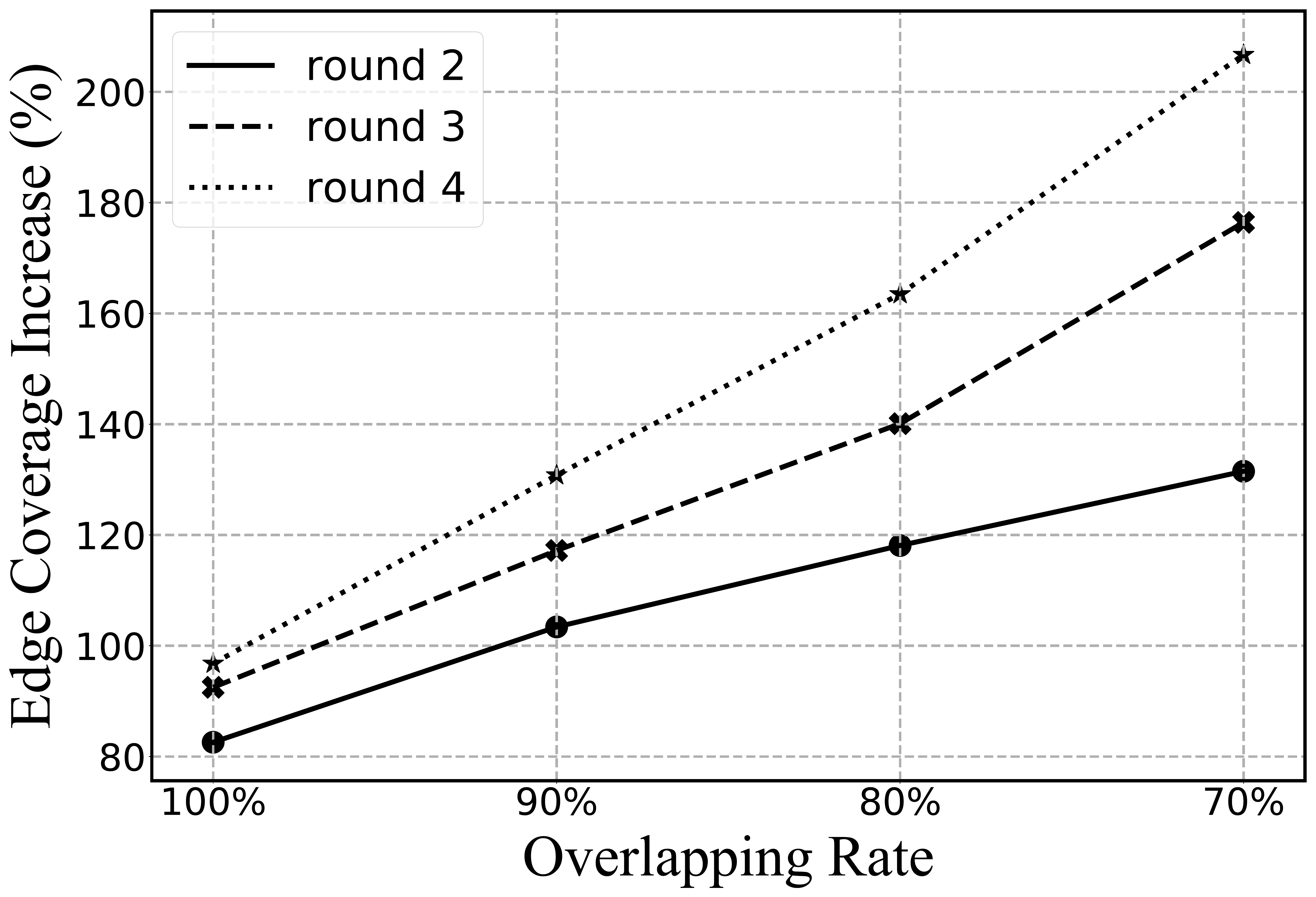} &
        \includegraphics[width=0.24\linewidth]{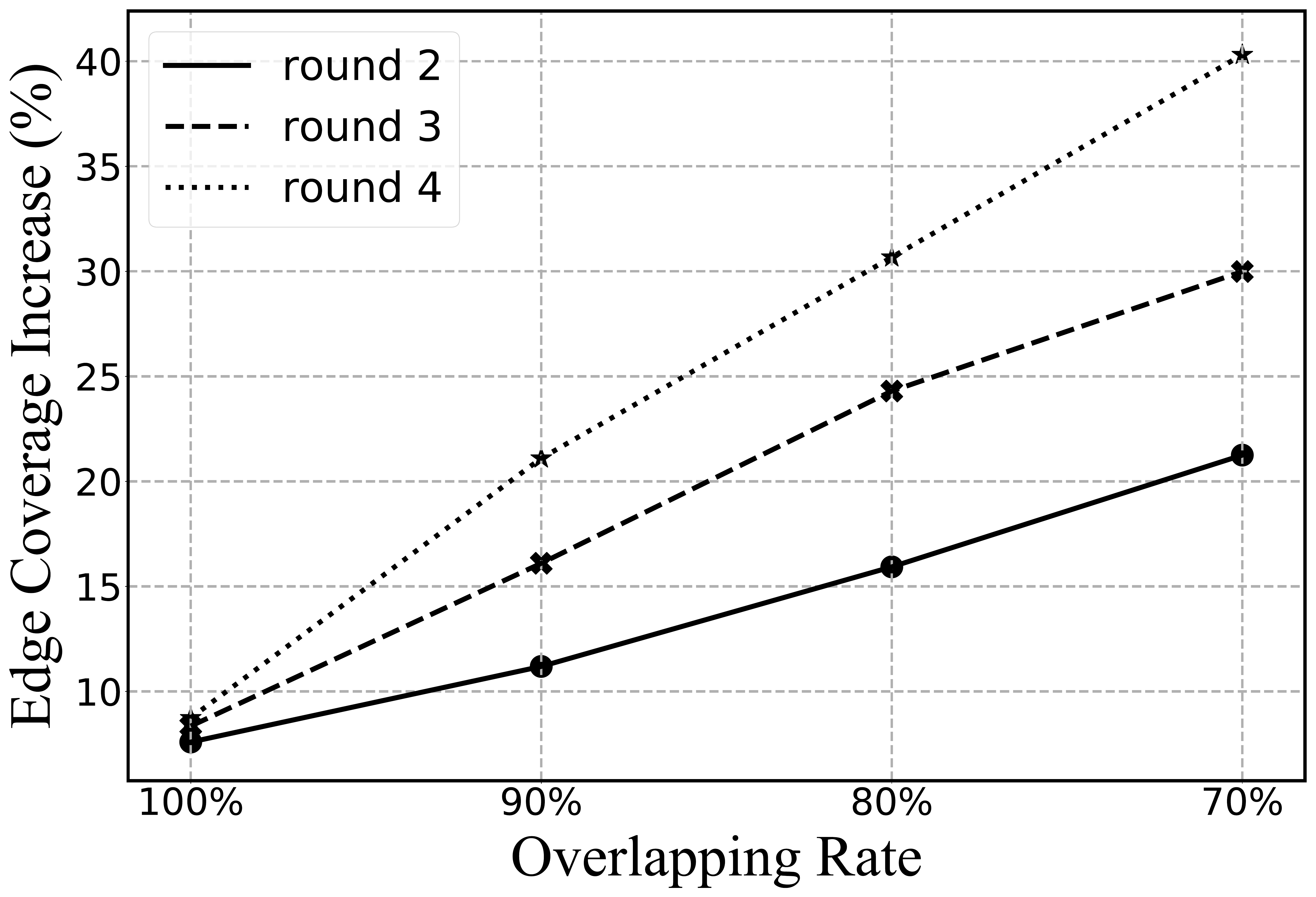} & 
        \includegraphics[width=0.24\linewidth]{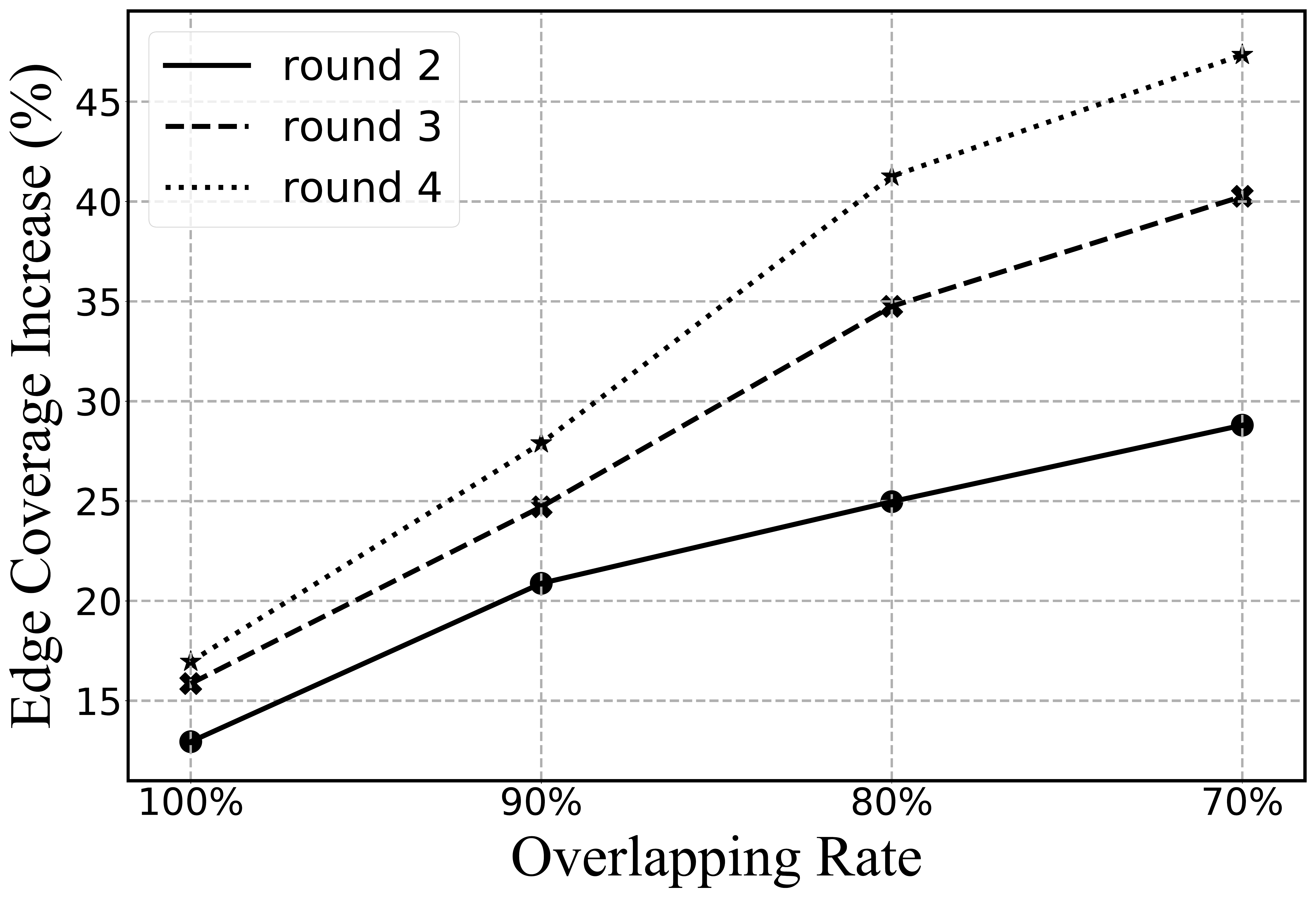} \\
        \centering{\hspace{0.6cm}(a) \sc objdump} & {\hspace{0.7cm}(b) \sc readelf} &
        \centering{\hspace{0.6cm}(c) \sc libxml} & {\hspace{0.6cm}(d) \sc tiff2ps} \\
    \end{tabular}
    \vspace{0.5em}
\caption{Impacts of overlapped mutations to edge coverage. The baseline of ``Edge Coverage Increase'' is the edges covered by the original 1,000 seeds; ``Overlapping Rate'' indicates the portion of overlapped seeds between two consecutive rounds of mutations. Results from the first round are omitted since the first round is identical under different settings, \ie mutating the initial 500 seeds. \emph{Please note that the x-axis decreases from left to right.}}
\label{fig:study-code-cov}
\vspace{-1em}
\end{figure}

Fig.~\ref{fig:study-code-cov} presents the results of the above experiment under different settings. With all the four programs, we observe a trend that fewer overlaps among the mutated seeds lead to a higher increase of code coverage. This empirically demonstrates that running \afl's mutations on different seeds can cover more edges than running the mutations on the same seeds. We believe such results align with \afl's design: \afl distributes mutation energies in a round according to the potential of a seed (based on metrics such as number of edges covered by the seed) and assigns more mutation cycles to seeds with higher potential, which eventually helps allocate sufficient mutations in a single round to exhaust the edges that can be derived from a seed.

To sum up, our empirical study shows that the parallel mode of \afl (and likely, many other fuzzing tools) can indeed bring overlaps, which further impedes the efficiency of code coverage. It is, therefore, necessary to investigate and develop better solutions of parallel fuzzing.

\section{A General Model of Parallel Fuzzing}
\label{sec:model}

In this section, we propose a model to describe parallel fuzzing. The model is inclusive of the parallel mode in existing fuzzing tools and we envision it is general enough to apply to other solutions of parallel fuzzing.

Formally, a parallel fuzzing system consists of $n$ instances \{$F_1$, $F_2$, ..., $F_n$\}. These instances together work on a set of $m$ tasks \{$T_1$, $T_2$, ..., $T_m$\}, with the $i^{th}$ instance distributed to focus on a subset of tasks \{$T_{i1}$, $T_{i2}$, ..., $T_{im_{i}}$\} ($m_{i}\le m$). Depending on the definition of tasks, \{$T_1$, $T_2$, ..., $T_m$\} can require different amounts of workload to be completed, notated as \{$W_1$, $W_2$, ..., $W_m$\}. To improve the efficiency of parallel fuzzing, the system would desire to meet the following three properties.

\begin{ditemize}
\item ($\mathbb{P}_1$) Different instances should work on disjoint subsets of tasks. This is to avoid overlaps and increase the extent of concurrency. Formally, given any two instances $F_i$ and $F_j$ ($i\ne j$), the fuzzing system needs to ensure:

\vspace{-1.2em}
\begin{equation}
\begin{aligned}
\label{eqn:no-overlap}
\{T_{i1}, T_{i2}, ..., T_{im_{i}}\} \cap \{T_{j1}, T_{j2}, ..., T_{jm_{j}}\} = \emptyset \end{aligned}
\end{equation}

\item ($\mathbb{P}_2$) All the instances together should cover all the tasks. Formally, that means:

\vspace{-1.2em}
\begin{equation}
\begin{aligned}
\label{eqn:coverall}
\bigcup_{i=1}^{n} \{T_{i1}, T_{i2}, ..., T_{im_{i}}\} = \{T_1, T_2, ..., T_m\}
\end{aligned}
\end{equation}

Otherwise, the pursuit of parallel fuzzing can cause the loss of certain tasks and, essentially, miss the code that can be covered by those tasks. 

\item ($\mathbb{P}_3$) Different tasks should be assigned with a similar workload. Formally, the fuzzing system should maintain the following relation between any two instances $F_i$ and $F_j$ ($i\ne j$):

\vspace{-1.2em}
\begin{equation}
\begin{aligned}
\label{eqn:workload}
\sum_{n=1}^{m_i} W_{in} \approx \sum_{n=1}^{m_j} W_{jn}
\end{aligned}
\end{equation}

Otherwise, certain instances can receive under-loaded tasks and end with plenty of idle cycles, which in principle also harms concurrency. 
\end{ditemize}

While the above model is general, it overlooks the fact that different tasks can bring different benefits, in particular code coverage~\cite{chen2019savior}. To this end, Equation~\ref{eqn:no-overlap} should be amended to allow overlaps of tasks that carry higher returns such that these tasks have a higher chance to be picked and completed. To incorporate this consideration, we find that it is not mandatory to modify our model. Instead, we can replicate high-return tasks and consider their replicas as unique ones. This will achieve similar effects as allowing overlaps of those tasks.

\section{Applications of Our Model}
\label{sec:app}

\subsection{Existing Solutions}

\begin{wrapfigure}{r}{0.5\textwidth}
\vspace{-2.5em}
\begin{center}
    \centering
    \includegraphics[width=1\linewidth]{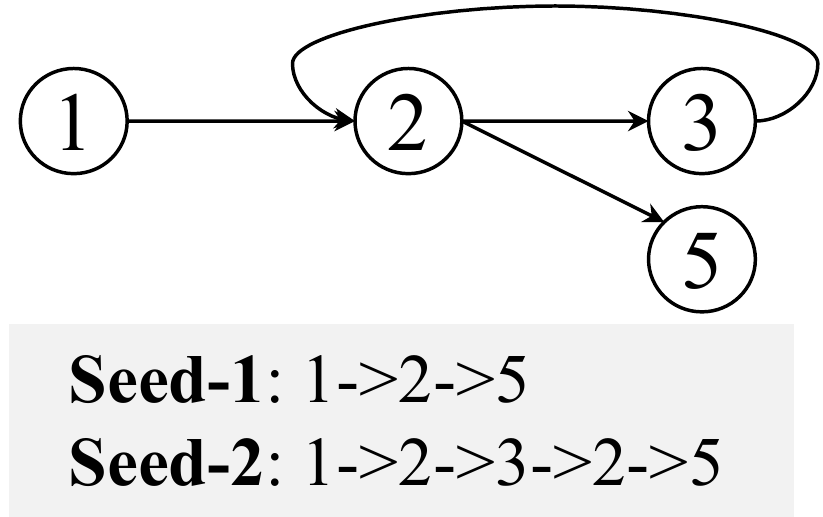}
    \caption{An example of two seeds that cover overlapped sets of edges. The \emph{upper} part presents the CFG; the \emph{lower} part shows the two seeds.}
    \label{fig:aflpaths}
    \vspace{-1em}
\end{center}
\vspace{-1em}
\end{wrapfigure}

In the literature, two solutions of improving parallel fuzzing, \pfuzz~\cite{song2019p} and \pafl~\cite{liang2018pafl}, fit into our model. Both \pfuzz and \pafl consider \emph{a fuzzing task is to run a round of mutations to a unique seed} and they distribute a similar amount of unique seeds to each instance. However, the two solutions take two opposite principles to define the uniqueness of a seed. \pfuzz aims for conservativeness. It follows \afl and considers a seed that brings new code coverage at its birth to be unique. Such a principle preserves all seeds produced by \afl but can leave behind many overlaps. Consider Fig.~\ref{fig:aflpaths}, where {\tt Seed-1} is born before {\tt Seed-2}, as an example. \pfuzz will consider both seeds unique and mutate both seeds. However, {\tt Seed-2} covers all the edges of {\tt Seed-1} and thus, they actually overlap based on \afl's definition. Moreover, mutating {\tt Seed-2} can likely produce the same code coverage as mutating both seeds, further illustrating the overlap. 

In contrast, \pafl aims for effectiveness. It considers a seed unique only when the seed covers certain less-frequently visited edges. In this way, \pafl can massively reduce overlaps. However, it does not guarantee that the unique seeds can cover all the code covered by the original seeds. As such, it can skip certain code regions and lose the opportunities to cover code that can be derived from those code regions.


\begin{figure}[tb]
    \centering
    \includegraphics[width=\linewidth]{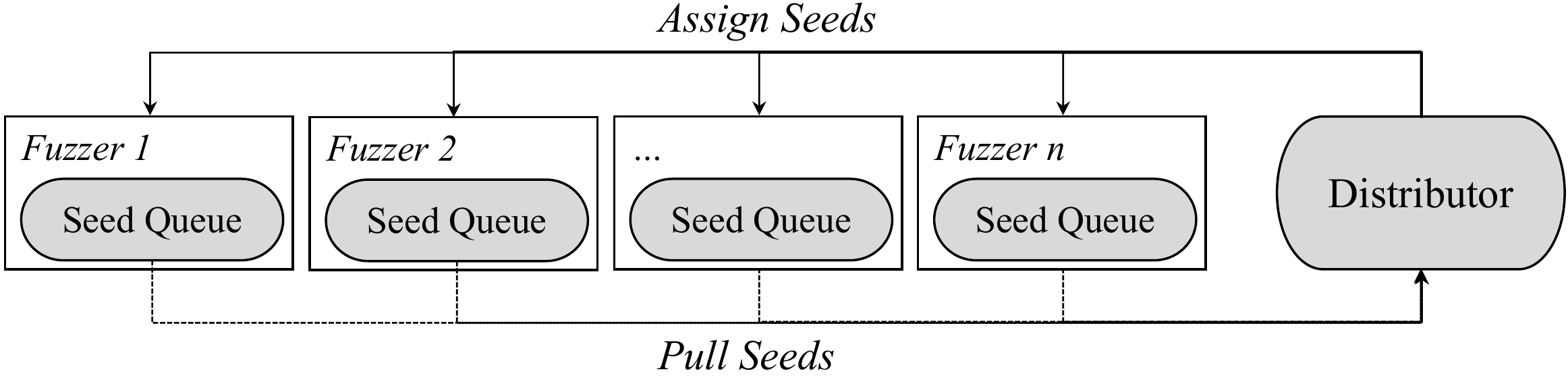}
    \caption{Workflow of our solution to optimize parallel fuzzing.} 
    \label{fig:workflow}
    \vspace{-1em}
\end{figure}

\subsection{Our Solution}
\label{subsec:codepartition}

In this paper, we propose a new solution following our model. Similar to \pfuzz and \pafl, we consider a fuzzing task is to run a round of mutations to a unique seed. However, as we will explain shortly, we adopt a strategy that achieves an effectiveness-and-conservativeness balance to define the uniqueness of a seed. Our solution follows the workflow in Fig.~\ref{fig:workflow} to periodically distribute the tasks. In each round of task distribution, we pull seeds from all instances, partition them into un-overlapped while similarly-weighted sub-sets, and finally assign them back to each instance. In the rest of this section, we describe the design details and explain how they meet $\mathbb{P}_1$ - $\mathbb{P}_3$. 

\noindent\textbf{Defining fuzzing tasks.} In our solution, we consider the entire set of tasks are to mutate seeds that cover all the (control flow) edges reached by the original seeds. In this way, we approximate the fuzzing goals of \afl and largely preserve the fuzzing space of \afl (\ie produce similar effects as mutating every seed). To determine the uniqueness of a seed, we consider the edges and their hit counts\footnote{\label{note1}A hit count in each of the following ranges is mapped to a unique value: \texttt{[1], [2], [3], [4, 7], [8, 15], [16, 31], [32, 127], [128, $\infty$)}} covered by the seed as criteria. But different from \pfuzz, we consider \emph{a seed is unique only if the seed covers one or more edges that other seeds do not cover}. This principle avoids the overlaps that \pfuzz may incur. Referring back to the example in Fig.~\ref{fig:aflpaths}, when {\tt Seed-1} and {\tt Seed-2} both exist at the moment of task distribution, we will consider {\tt Seed-1} non-unique since {\tt Seed-2} encapsulates all the edges of {\tt Seed-1}. In the following, we describe how to pick unique seeds while running task distribution. 

\begin{algorithm}[t!]
\caption{\textsc{Task distribution}}
\label{alg:taskdistribute}
\Input{Seed sets from all instances $\mathcal{D}=\{\vec{S}_1,\vec{S}_2, ..., \vec{S}_n\}$}
\Output{Seeds distributed to different instances $\mathcal{D'}=\{\vec{S}'_1,\vec{S}'_2, ..., \vec{S}'_n\}$}
Initialize $\mathcal{D'}$: $\vec{S}'_1=\emptyset$, $\vec{S}'_2=\emptyset$,... $\vec{S}'_n=\emptyset$

\For{each $\vec{S}_i\in\mathcal{D}$}{
    Obtain edges covered by seeds in $\vec{S}_i$, notated as $\vec{E}_i$\; \tcc{hit counts of edges are considered}
}
$\vec{E}=\bigcap_{i=1}^{n} \vec{E}_{i}$\;

Organize $\vec{E}$ into a control flow graph $CFG$\; \tcc{different hit counts of the same edge are represented as different edges}

Copy $CFG$ as $CFG'$ and topologically sort $CFG'$\;

\For{the deepest leaf node $L_i$ in $CFG'$}{
    $k=random(1, n)$\;
    
    Pick a seed $s$ from $\vec{S}_k$ which covers $L_i$, maximizes $|edge(s) \bigcap edge(CFG')|$, and has the minimal age\;
    
    Add $s$ to  $\vec{S}'_k$\;
    
    Remove $edge(s)$ from $CFG'$\;
}

\For{each $\vec{S}_i\in\mathcal{D}$}{
    \For{each $s\in\vec{S}_i$}{
        \If{$edge(s)-edge(CFG)\neq\emptyset$ and $edge(s)-edge(\vec{S}'_i)\neq\emptyset$}{
            Add $s$ to $\vec{S}'_i$\;
        }
    }
}
\Return{$\mathcal{D}'$}\;
\end{algorithm}

\begin{figure}[tb]
    \centering
    \includegraphics[width=\linewidth]{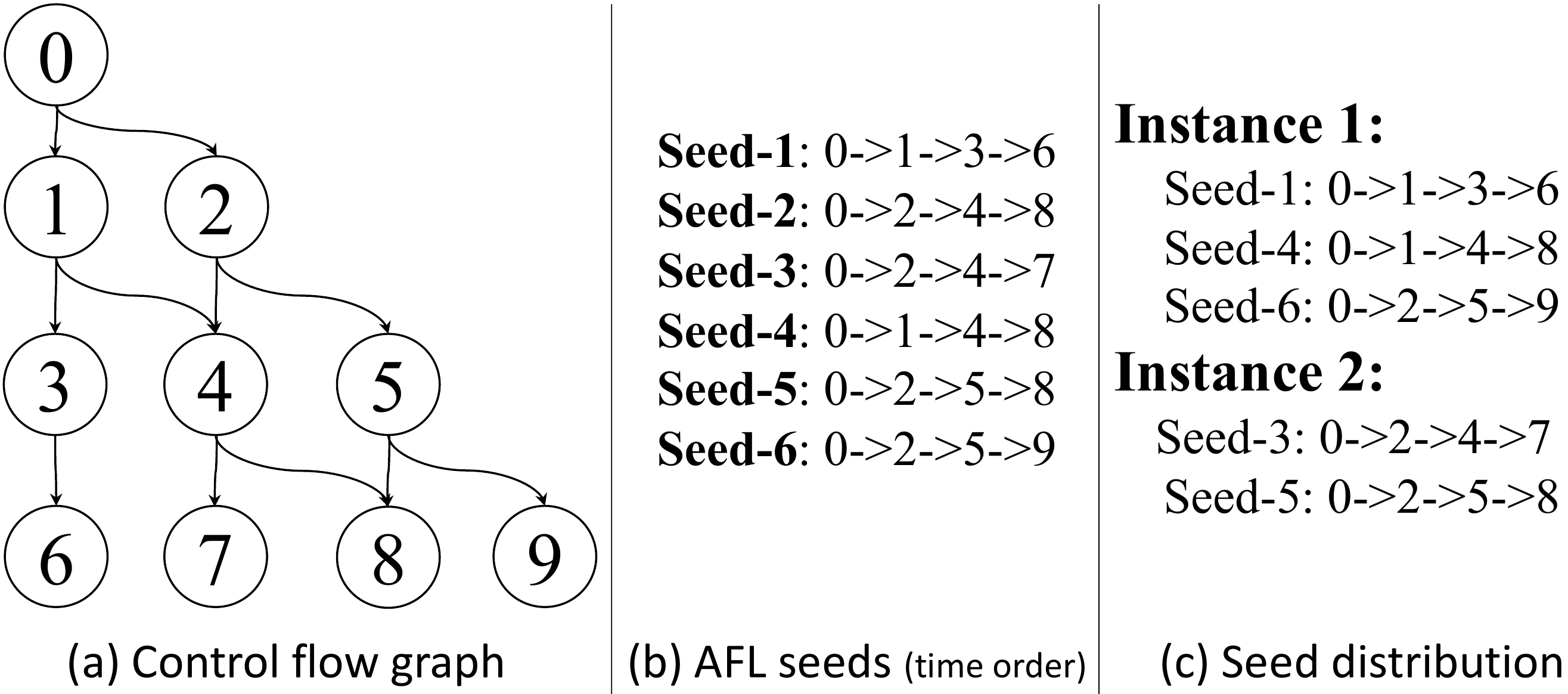}
    \caption{An example of edge-coverage-based task distribution between 2 instances. The \emph{left} part shows the CFG aggregated by the overlapped edges.  The \emph{middle} part show the seeds produced by \afl, sorted in time order. The \emph{right} part presents the task distribution results.}
    \label{fig:edge-split}
    \vspace{-1em}
\end{figure}

\noindent\textbf{Distributing fuzzing tasks.} In each round of task distribution, we pull all the seeds from each instance and re-run them with dynamic tracing. We gather the edges covered by each instance, notated as $\vec{E}_{i}$ for the $i^{th}$ instance. We then compute the intersections among the edges from all instances (\ie $\bigcap_{i=1}^{n} \vec{E}_{i}$), and we notate the intersections as $\vec{E}$. By intuition, we can then randomly, evenly partition $\vec{E}$ into multiple sub-sets, assign each sub-set to a unique instance, and then pick seeds that visit those assigned edges for the instance to mutate. Such an idea, however, has a major problem. When we pick a seed to cover a particular edge, we will concurrently cover many other edges, which may essentially bring overlaps back. Consider the Fig.~\ref{fig:edge-split} as an example. By random distribution, we may sequentially pick {\tt Seed-2}, {\tt Seed-3}, and {\tt Seed-4}, and distribute them to different instances. According to our definition, this would create an edge-overlap between {\tt Seed-2} and {\tt Seed-3 + Seed-4}, not satisfying our definition of uniqueness. 

In this work, we design a greedy algorithm, shown in Algorithm~\ref{alg:taskdistribute}, to provide \emph{edge-coverage-based task distribution}. We aggregate the edges in $\vec{E}$ into a topologically sorted control flow graph, notated as $CFG$ (line {\sf 1-6}). We then recursively process the leaf edges on $CFG$ (i.e., edges that end with leaf nodes on $CFG$). For the leaf edge with the largest depth, we randomly pick a fuzzing instance and elaborately pick a seed $s$ from that instance to cover the leaf edge (line {\sf 9-10}). To be specific, we select the seed that covers the maximal number of edges remaining on the $CFG$. If multiple seeds satisfy this condition, we pick the youngest one. We distribute $s$ to the fuzzing instance where $s$ comes from (line {\sf 11}) and remove all edges covered by $s$ from $CFG$ (line {\sf 12}). We repeat this process until all edges on the $CFG$ are removed. After that, we preserve the seeds that visit edges in $\neg \vec{E}$ (line {\sf 14-20}). To better illustrate our distribution algorithm, we present an example in Fig.~\ref{fig:edge-split}, showing both the fuzzing progress and the distribution results. It is worth noting that when we pick a seed for leaf node {\tt 4$\to$8}, we favor {\tt Seed-4} over {\tt Seed-2} because  {\tt Seed-4} is more recently derived. This prevents the pick of {\tt Seed-2} and avoids the overlap we mentioned before.

The above algorithm involves multiple heuristics, which strive for fewer overlaps and better efficiency. First, we prefer seeds that cover more non-distributed edges (line {\sf 10}). The motivation is to quickly consume the distribution space and thus, minimize the number of required seeds and reduce the potential of overlaps. Second, we favor newer seeds. The rationale is that seeds newly generated have a higher chance to cover new edges than the older seeds. Thus, they have a lower chance of bringing in overlaps. Third, we prioritize edges that have larger depths on the control flow graph. This is to reduce the search space when picking seeds, exploiting the observation that deeper edges are typically reached by fewer seeds. 

Despite our greedy algorithm may not perfectly meet $\mathbb{P}_1$ - $\mathbb{P}_3$, it represents the best effort. First, we distribute disjoint sub-sets of overlapped edges to different instances. We further incorporate a set of heuristics to avoid overlaps when we pick seeds. As we will demonstrate in \S~\ref{sec:eval}, this combined effort can indeed effectively reduce overlaps (way more than \pfuzz). Second, every edge (in $\vec{E}$ or $\neg \vec{E}$) is distributed to at least one instance. This preserves all the fuzzing tasks according to our definition, satisfying $\mathbb{P}_2$. Finally, we evenly distribute the non-overlapped edges in a random manner. This aids each instance to receive approximately equivalent workloads and, therefore, facilities the fulfillment of $\mathbb{P}_3$.
 
\noindent\textbf{Scheduling task distribution.} Our design needs to periodically re-run the task distribution. However, a low frequency of re-distribution may not timely avoid the accumulated overlaps while a high frequency can lead to a waste of computation cycles since fuzzing may not have produced many overlaps. In our design, we adjust the scheduling of task distribution based on the increase of edge coverage. We start the first round of distribution after the first hour, and we re-run it once the new edge coverage exceeds 10\%.

\section{Implementation} 
\label{sec:impl}

We have implemented our solution, called \afledge, on top of \afl (2.52b) and LLVM with around 100 lines of C code, 400 lines of C++ code, 300 lines of shell scrips, and 200 lines of Python code. All code will be released upon publication.

\subsection{Collecting Edge Coverage.} The task distribution of \afledge needs code coverage information of existing seeds. To support the need, we implement an LLVM pass to instrument the target program. Following a seed, the instrumented code will sequentially record each edge and output the final list at the end. To avoid collisions, we assign each basic block a unique 64-bit ID and concatenate the IDs of two connected basic blocks to represent the edge between them.

\subsection{Distributing Fuzzing Tasks.} \afledge requires to distribute seeds across fuzzing instances. To avoid intruding on the normal fuzzing process, we implement the task distributor as a standalone component. It follows the algorithm in \S~\ref{subsec:codepartition} to determine the seeds that are assigned to each instance and saves the seeds in a file. Following the metadata organization of \afl, the seed file is added to the corresponding instance's working directory. 

\subsection{Confining Fuzzing Tasks.} Our design requires an instance to only mutate the sub-group of assigned seeds. Technically, we customize \afl to read the list of seeds assigned by the distributor and maintain them in a allow-list. When \afl schedules seeds for mutations, we only pick candidates on the allow-list. Such an implementation avoids introducing extra inconsistency to the fuzzing process. Considering that our distributor iteratively updates the seed list, the customized \afl periodically checks the seed file and updates the allow-list accordingly.

\section{Evaluation}
\label{sec:eval}

In this section, we evaluate \afledge, centering around three questions. 
\begin{ditemize}
\item \emph{($\mathbb{Q}_1$) Can \afledge reduce the overlaps among fuzzing instances?}
\item \emph{($\mathbb{Q}_2$) Can \afledge improve the efficiency of code coverage?}
\item \emph{($\mathbb{Q}_3$) Can \afledge preserve the fuzzing capacity of \afl?}
\end{ditemize}

\subsection{Experimental Setup}
\label{subsec:eval:setup}

\begin{table}[!t]
\centering
\caption{Benchmark programs and evaluation settings. 
In the column of {\tt Seeds}, {\tt AFL} means we reuse the test-cases from \afl and {\tt built-in} means that we reuse the test cases from the program.}
\begin{tabular}{lccc||cc}
\toprule[0.5pt]
\toprule[0.5pt]

\multicolumn{4}{c||}{\bf{\emph{Programs}}} & \multicolumn{2}{c}{\bf{\emph{Settings}}}   
 \\ \hline
 {\tt Name} & {\tt Version} & {\tt Driver} & {\tt Source} & {\tt Seeds} & {\tt  Options}
 \\ \hline

{\sc libpcap} & 1.10.0 & {\sc tcpdump} & ~\cite{Indexofr65:online} & \afl & -r @@ \\
{\sc libtiff} & 4.0.10 & {\sc tiff2ps} & ~\cite{Indexofl62:online} & \afl &  @@ \\

{\sc libtiff} & 4.0.10 & {\sc tiff2pdf} & ~\cite{Indexofl62:online} & \afl &  @@ \\
{\sc binutils} & 2.32 & {\sc objdump} & ~\cite{Indexofg63:online} & \afl &  -d @@ \\

{\sc binutils} & 2.32 & {\sc readelf} & ~\cite{Indexofg63:online} & \afl &  -a @@ \\
{\sc binutils} & 2.32 & {\sc nm-new} & ~\cite{Indexofg63:online} & \afl &  -a @@ \\

{\sc libxml2} & 2.9.7 &  {\sc xmllint} & ~\cite{libxml2src} & \afl &  @@ \\
{\sc nasm} & 2.14.2 & {\sc nasm} & ~\cite{nasmsrc} & built-in &  -e @@ \\

{\sc ffmpeg} & 4.1.1 & {\sc ffmpeg} & ~\cite{ffmpegsrc} & built-in &  -i @@ \\
\bottomrule[0.5pt]
\bottomrule[0.5pt]
\end{tabular}
\label{tab:eval-setup}
\end{table}

\noindent\textbf{Benchmarks.} To answer the above questions, we prepare a group of 9 real-world benchmark programs. Details about the programs are presented in Table~\ref{tab:eval-setup}. All these programs have been intensively tested in both industry~\cite{ossfuzz} and academia~\cite{stephens2016driller,qsyminsu,vuzzer}. In addition, they carry diversities in both functionality and complexity.

\noindent\textbf{Baselines.} We run \afl as the baseline of our evaluation. 
To compare \afledge with the existing solutions, we also run \pfuzz~\cite{song2019p} and \pafl~\cite{liang2018pafl} on top of \afl 
. Because the implementations of \pfuzz and \pafl are not publicly available, we re-implemented the two solutions following the algorithms presented in their publications~\cite{song2019p,liang2018pafl}.

\noindent\textbf{Configurations.} Specific configurations of the fuzzing process (\eg seeds and program options) are listed in Table~\ref{tab:eval-setup}. To understand the impacts of the number of instances, we run each fuzzing setting respectively with 2, 4, and 8 \afl secondary instances. We do not run a primary instance because it involves deterministic mutations which bring disadvantages to vanilla \afl. For consistency, we conduct all the experiments on Amazon EC2 instances (Intel Xeon E5 Broadwell 96 cores, 186GB RAM, and Ubuntu 18.04 LTS), and we sequentially run all the tests to avoid interference. Each test is run for 24 hours. To minimize the effect of randomness in fuzzing, we repeat each test 5 times and report the average results.


\subsection{Analysis of Results}
\label{subsec:eval:result}

In Table~\ref{tab:afl:stat}, we present the results with \afl at the end of 24 hours. We elaborate on the results as follows, seeking answers to $\mathbb{Q}_1$ - $\mathbb{Q}_3$. 

\noindent\textbf{Effectiveness of overlap reduction.} The direct goal of \afledge is to reduce the overlaps among instances. To measure this goal, we consider the number of seeds that are disabled from each instance as the metric. As shown in Table~\ref{tab:afl:stat} (the column for \textit{overlap reduction rate}), \afledge can effectively reduce the potential overlaps in the parallel mode of \afl. To be specific, \afledge can prevent 60.0\%, 60.3\%, and 57.1\% of the seeds from being repeatedly mutated when we respectively run 2, 4, and 8 parallel instances.  

\begin{table*}[t!]
\scriptsize
\caption{Statistical results of our evaluation with \afl in 24 hours. In the table, ``\textsf{overlap reduction (\%)}'' means the average percentage of seeds that the corresponding solution cuts from each instance; ``\textsf{edge-cov increase (\%)}'' stands for the increase of code coverage that the corresponding solution brings to \afl; \textsf{p-value} demonstrates the statistical significance of the increase of code coverage (the smaller, the better); and ``\textsf{edge-cov overlap rate (\%)}'' shows how much of the code covered by \afl is also covered by the corresponding solution.}
\label{tab:afl:stat}
\vspace{0.5em}
\begin{tabular}{p{1cm}|wl{1.2cm}||p{0.70cm}p{0.70cm}p{0.70cm}|p{0.70cm}p{0.70cm}p{0.70cm}|p{0.70cm}p{0.70cm}p{0.70cm}|p{0.70cm}p{0.70cm}p{0.70cm}}
\toprule[0.5pt]
\toprule[0.5pt]
\multirow{2} {*}{\textbf{\normalsize{Prog.}}} & 
\multirow{2} {*}{\textbf{\normalsize{Tool}}} &
\multicolumn{12}{c}{\textbf{Statistical evaluation results with 2, 4, and 8 instances (\afl).}} \\ \cline{3-14}
& & \multicolumn{3}{c|}{\scriptsize \textsf{overlap reduction (\%)}} & \multicolumn{3}{c|}{\scriptsize \textsf{edge-cov increase (\%)}} & \multicolumn{3}{c|}{\scriptsize \textsf{p-value}} & \multicolumn{3}{c}{\scriptsize \textsf{edge-cov overlap (\%)}} \\

\hline
\multirow{3}{*}{\sc objdump} &{\sc pafl}  &60.6 & 80.8 & 89.5            &1.9 &0.0  & 0.3                 &0.32    &0.72    &0.10      &95.5 &94.9 &95.8
\\
& {\sc p-fuzz}  &46.7 &53.5 &62.5            &5.1 & 2.3 & 3.0                & 0.07   & 0.03   & 0.00      &96.8 &96.3 &97.9 \\

& {\sc afl-edge} & 63.7 & 63.2 & 62.4             &7.6 & 7.3 & 3.1         & 0.00 & 0.04 & 0.03                       &97.5 & 98.3 & 98.5 \\

\hline

\multirow{3}{*}{\sc readelf} &{\sc pafl}        &81.2 &86.8 &88.4          &5.6 &-3.2 &-3.6        &0.03  &0.19  &0.12      & 92.5 &92.1 &93.0                                      \\

&{\sc p-fuzz}        &46.8 & 43.8 & 45.4          &7.2 &3.8 &6.0        &0.04  &0.01  &0.00             &99.0 &98.5 &98.0                                     \\

&{\sc afl-edge}  &45.9 & 43.8 & 42.5          &12.2 &7.0 &8.4        &0.05  &0.02  & 0.00       &97.5 &97.2 &97.7                                     \\

\hline

\multirow{3}{*}{\sc tiff2pdf} &{\sc pafl}  &64.1 &79.7 &81.2           &3.4 &0.4 &5.0           &0.00 &0.05 &0.29      &97.0 &91.3 &93.8                                      \\

&{\sc p-fuzz}       &42.2 & 60.1 & 64.1          &5.5 &6.3 &8.5          &0.01 &0.04 &0.04               &98.5 &97.0 &97.9                                     \\

&{\sc afl-edge}    &62.4& 59.5& 58.5          &9.0& 5.6& 11.2        & 0.00 & 0.02 &0.00       &98.5& 95.9& 99.1                   \\

\hline
\multirow{3}{*}{\sc nm-new} &{\sc pafl}    &46.2 &52.6 &55.9                         &4.0 &4.8  &0.2             &0.10  &0.07  & 0.19                &96.6 &96.2  &97.5                                       \\

&{\sc p-fuzz}    &43.6& 55.6& 60.2                        &4.5& 7.1& 3.8           & 0.00 &0.03 &0.00               &97.6& 98.1& 98.1                                     \\

 &{\sc afl-edge}      &59.9& 58.5& 60.1                        &6.8  &6.6 &3.6         &0.03 &0.08 &0.00               &96.3& 96.8& 98.5        \\

\hline

\multirow{3}{*}{\sc nasm} &{\sc pafl}        &64.2 &82.7  &54.7          &5.5 &14.6  &2.4             &0.05   &0.01  &0.06               &99.8 &99.3  &99.8                               \\

&{\sc p-fuzz}        &10.0& 11.0 & 8.2        &8.9 & 9.2 & 8.1           & 0.00 & 0.00 & 0.00              &99.0& 98.3& 94.3                              \\

&{\sc afl-edge}    &67.9 & 68.3 & 66.8                        &13.9 & 22.6 & 7.6           & 0.00 & 0.00 &0.00              &99.2 & 99.3 & 99.2             \\

\hline

\multirow{3}{*}{\sc tiff2ps} &{\sc pafl}  &57.1 & 81.9 & 89.2       &7.8  &15.0  &8.7   &0.03 & 0.02 &0.02  &82.3 & 89.9 & 96.6  \\

&{\sc p-fuzz}  &44.0& 60.7& 66.4       &7.5 & 12.3 & 9.5   & 0.00 & 0.00 & 0.00  &97.1 & 97.5 & 98.0  \\

&{\sc afl-edge}  &53.4 &58.5  &52.8  &10.4& 13.5 & 7.2           & 0.00 &0.00 &0.00    & 96.9 & 96.5& 97.2          \\

\hline
\multirow{3}{*}{\sc tcpdump} &{\sc pafl}        &66.4 &80.3 &84.4        & 8.2 &10.6 & 7.7     &0.19 &0.03 &0.06       &84.7  &86.6 &88.5                                    \\

&{\sc p-fuzz}        &45.6& 63.6& 80.7         &2.8& 7.2& 6.1     &0.05 &0.03 &0.04       &85.7& 88.3& 91.5 \\

&{\sc afl-edge}       &48.3& 48.6& 42.9                    &4.4& 9.9& 11.9       &0.05 & 0.03 & 0.00      &87.6& 89.5& 92.8                                    \\

\hline
\multirow{3}{*}{\sc libxml2} &{\sc pafl}   &61.8 &91.5 &82.2     &7.3 &43.7 &7.0      & 0.00 &0.01 & 0.06      &87.2  &81.7 & 88.3                                      \\

&{\sc p-fuzz}   &41.0& 48.5& 51.8     &5.1& 11.7& 21.3      &  0.00 &0.01 &0.00      &97.9& 89.3& 97.6                                      \\

&{\sc afl-edge}   &68.9& 69.2& 65.6                       &7.5 & 7.6 & 29.1      & 0.00 &0.04 & 0.00                  &98.6 & 93.3 & 98.2              \\

\hline
\multirow{3}{*}{\sc ffmpeg} &{\sc pafl}       &94.5 & 90.7 & 91.4      &17.9 & 23.4 & 4.1     &0.04 &0.01 &0.38           & 86.8 & 85.2 & 87.0            \\

&{\sc p-fuzz}       &41.6& 62.5& 63.4      & 22.6 & 20.1 & 6.1     &0.03 &0.05 &0.00           & 91.2 & 90.2 & 98.3            \\

&{\sc afl-edge}   &69.7 & 73.5 & 62.3      &18.3 & 11.6 & 3.3      & 0.01 &0.04 &0.00              & 89.9 & 90.1 & 97.3               \\

\hline
\multirow{3}{*}{\textbf{Ave.}} & {\sc pafl} 
 &\textbf{{66.2}} &\textbf{80.8}&\textbf{79.7} &\textbf{6.9} &\textbf{12.1} &\textbf{3.5} &\textbf{--}&\textbf{--}&\textbf{--}  &\textbf{91.4}&\textbf{90.8}&\textbf{93.4}\\

& {\sc p-fuzz}  
 &\textbf{40.2} &\textbf{45.1}&\textbf{49.5} &\textbf{7.7} &\textbf{8.9} &\textbf{8.0} &\textbf{--}&\textbf{--}&\textbf{--}  &\textbf{95.9}&\textbf{94.8}&\textbf{96.9}\\

& {\sc afl-edge}
&\textbf{60.0}&\textbf{60.3}&\textbf{57.1} &\textbf{10.0} &\textbf{10.2}&\textbf{9.5} &\textbf{--}&\textbf{--} &\textbf{--}  &\textbf{95.8} &\textbf{95.2} &\textbf{97.6} \\

\toprule[0.5pt]
\toprule[0.5pt]
\end{tabular}
\end{table*}

In comparison to existing solutions, \afledge reduces more overlaps than \pfuzz but fewer than \pafl. Such results well comply with the designs of the three tools. \pfuzz preserves all the seeds produced by \afl while \pafl aggressively skip seeds. In contrast, \afledge keeps seeds necessary to cover all the edges, pursuing a trade-off between conservativeness and effectiveness. As we will show later, while \afledge's strategy reduces fewer seeds in comparison to \pafl, it does not necessarily hurt code coverage and it can better preserve the fuzzing capacity (or more precisely, \afledge can cover more code that \afl covers).


\noindent\textbf{Improvements to code coverage efficiency.} To understand whether the overlap reduction by \afledge can indeed benefit code coverage, we measure the number of edges covered in the tests. In Table~\ref{tab:afl:stat} (the column of \textit{edge coverage increase}), we present the increase of edge coverage brought by \afledge to \afl at the end of a 24-hour test. 

In summary, \afledge can consistently improve the efficiency of edge coverage of \afl, regardless of the benchmark and the number of instances. Specifically, \afledge increases the edge coverage by 10.0\%, 10.2\%, and 9.5\%, respectively with 2, 4, and 8 instances. Another key observation is that the benefits brought by \afledge often decrease with the number of instances. We believe the reason is that the fuzzers can get closer to saturation when more parallel instances are running. Therefore, the gap between \afl and \afledge shrinks at the end. 


To further verify that the improvements by \afledge are statistically significant, we perform Mann Whitney U-test~\cite{mcknight2010mann} on the five rounds of runs~\cite{evaluatefuzz}. The p-values of the hypothesis test are presented in Table~\ref{tab:afl:stat} (the column of \textit{p-value}). In nearly all the cases, the p-values are smaller than 0.05, supporting that the improvements brought by \afledge are significant from a statistical perspective.

Finally, \afledge presents better overall performance than both \pfuzz and \pafl. When applied to \afl, \afledge increases the edge coverage by 9.5\% - 10.2\%, outperforming \pfuzz and \pafl in most of the cases.
On the one hand, \afledge reduces more overlaps than \pfuzz and thus, produces higher code coverage efficiency. On the other hand, \pafl in principle reduces more overlaps than \afledge, which indeed leads to higher edge coverage than \afledge in several cases (\eg running \afl on {\sc TIFF2PS} with 4 or 8 instances). However, in many other cases, \pafl can accidentally block valuable seeds and become unable to cover the related edges, eventually resulting in a lower edge coverage. This can be further supported by that \pafl even produces lower edge coverage than \afl in certain cases (\eg running \afl on {\sc READELF} with 4 or 8 instances).


\begin{figure}[t!]
    \renewcommand{\arraystretch}{0.5}
    \begin{tabular}{cccc}
        \includegraphics[width=0.24\linewidth]{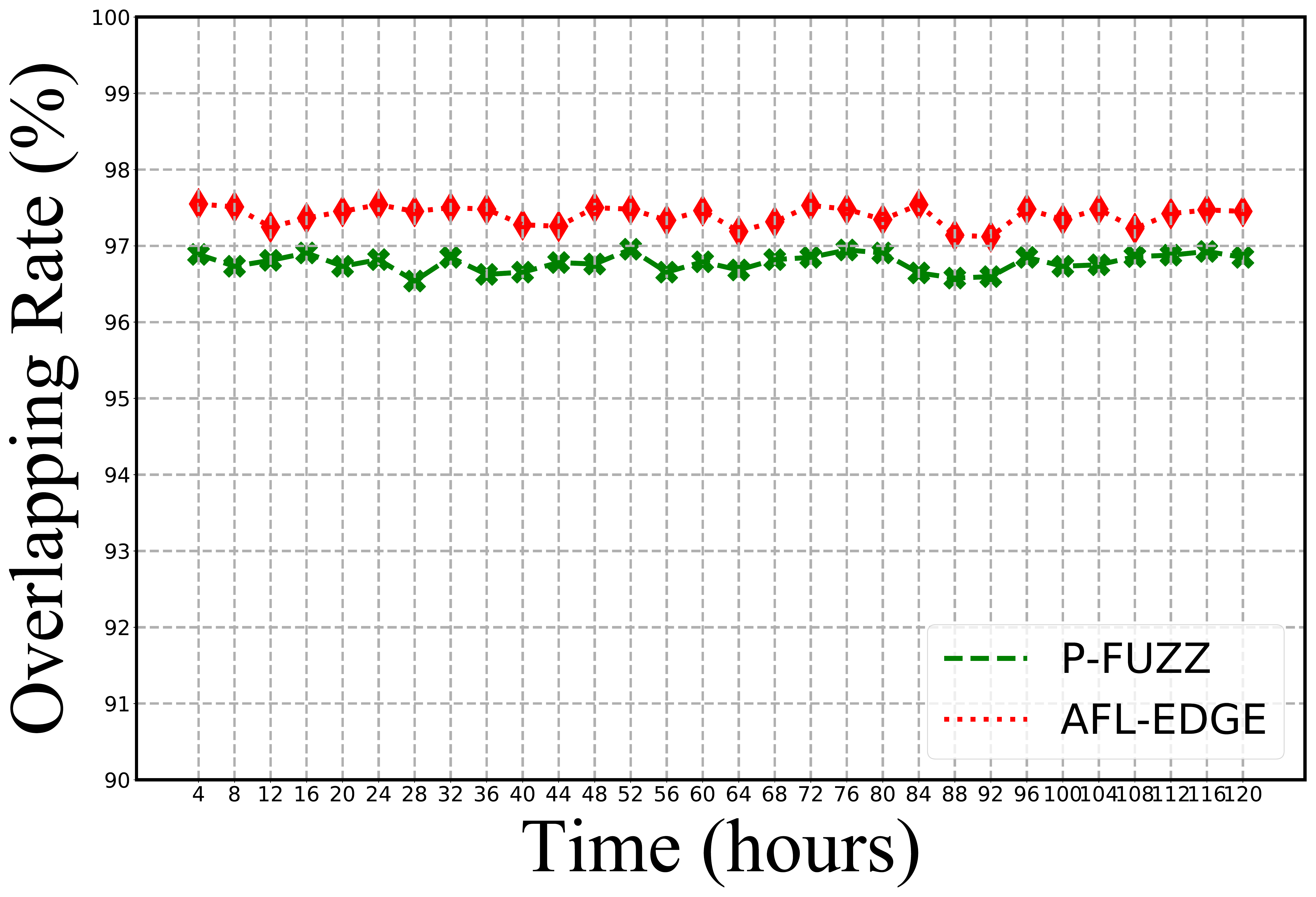} &
        \includegraphics[width=0.24\linewidth]{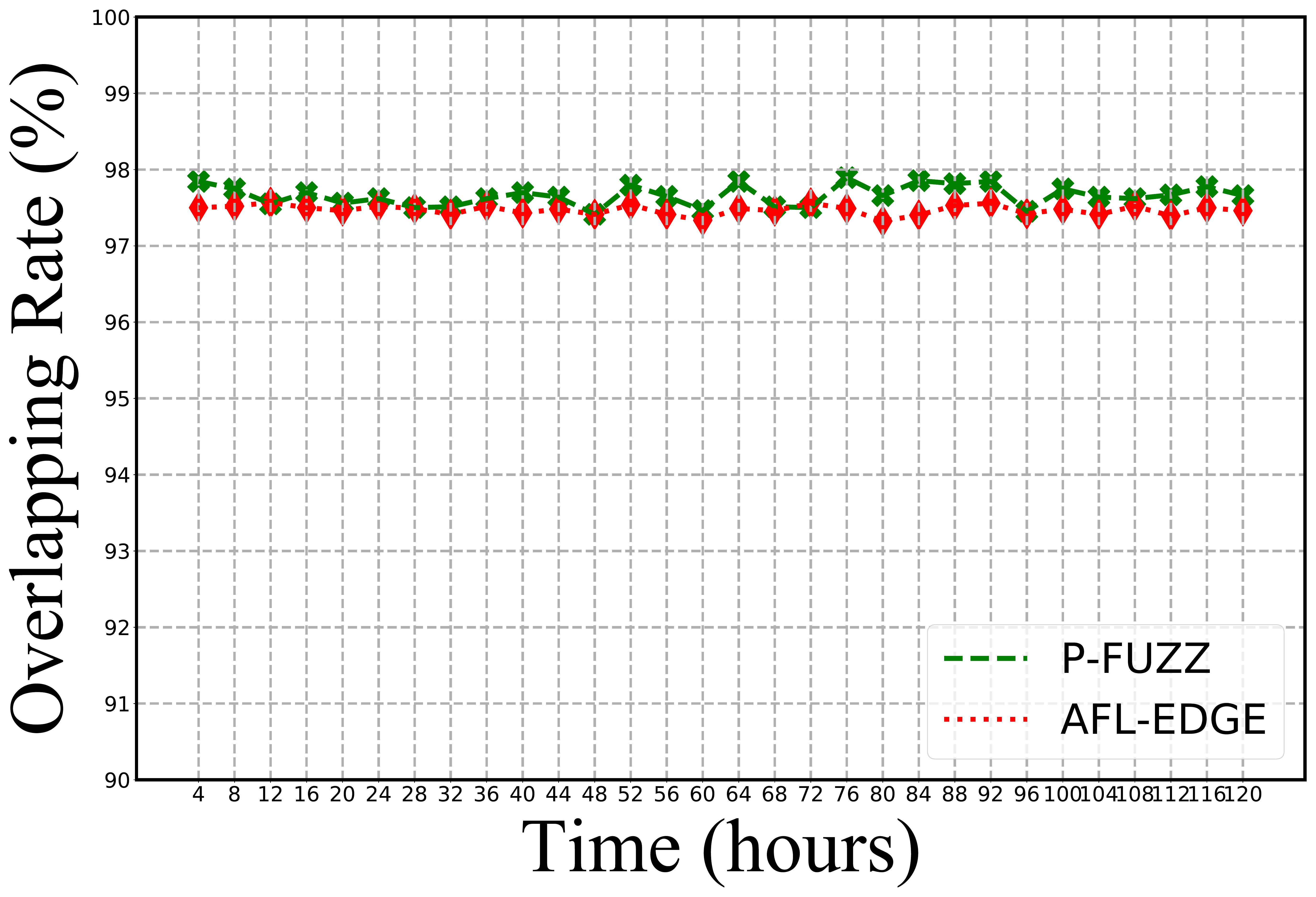}  &
        \includegraphics[width=0.24\linewidth]{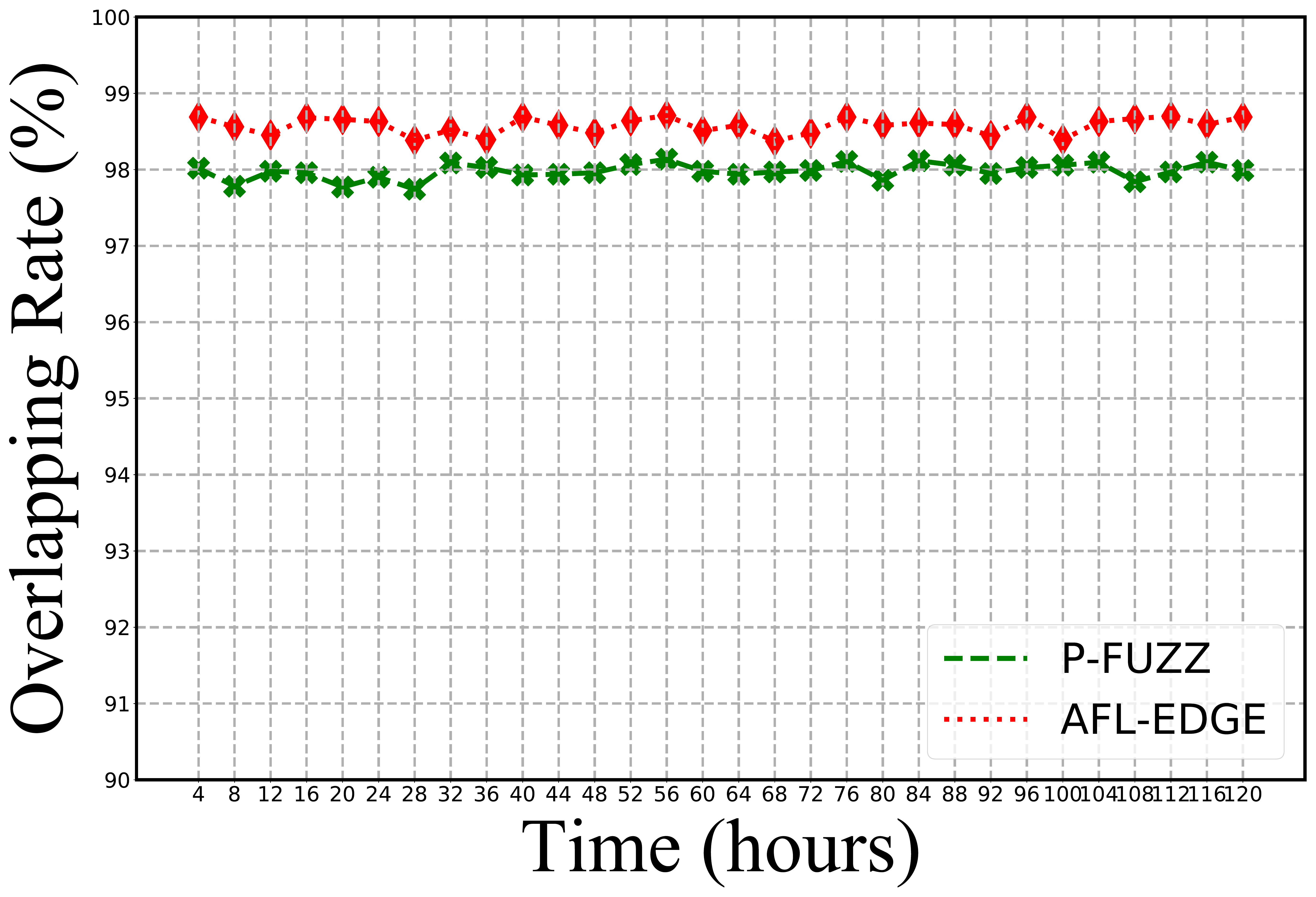} & \includegraphics[width=0.24\linewidth]{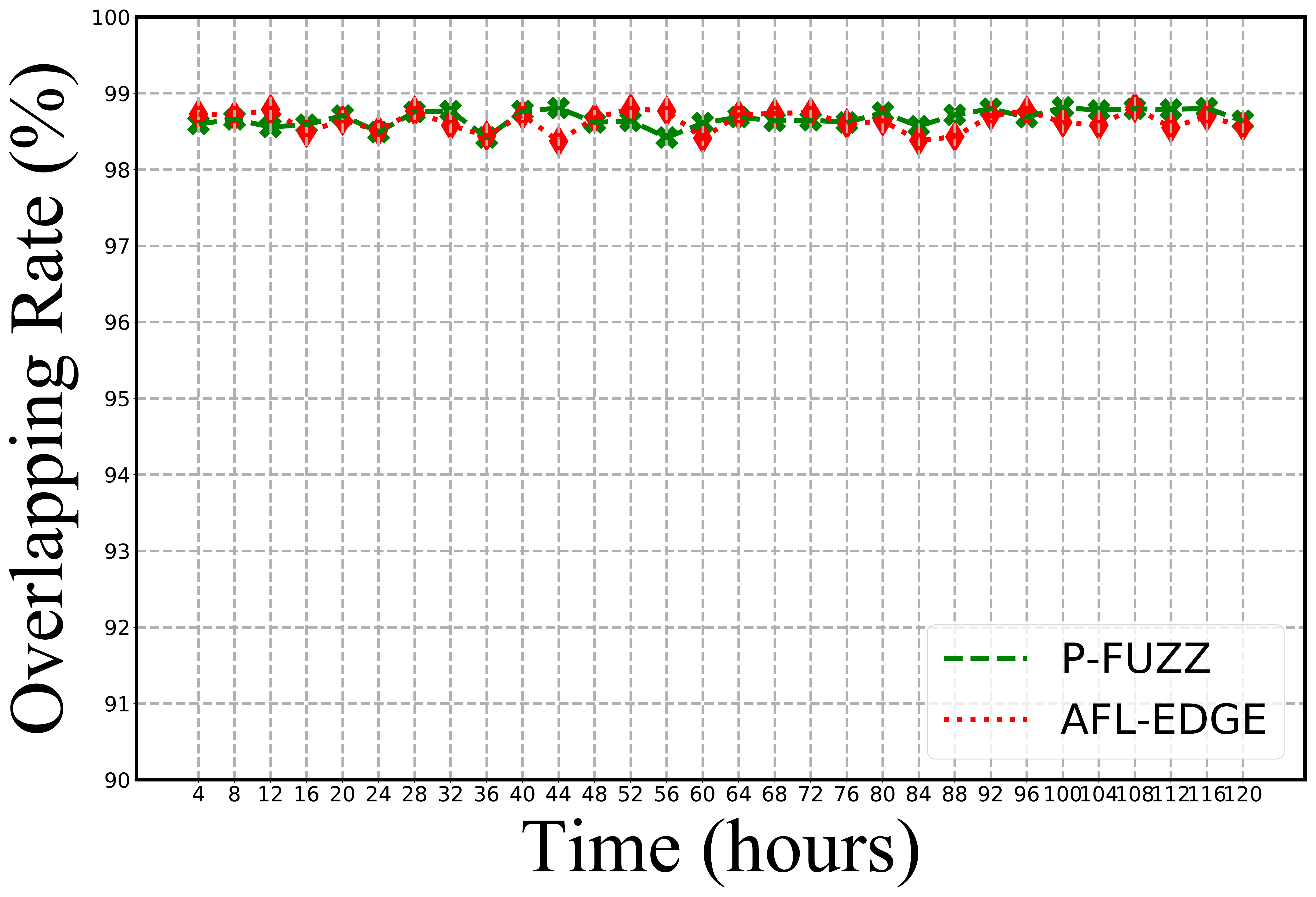} \\
        \centering{\hspace{0.6cm}(a) \sc objdump} & {\hspace{0.7cm}(b) \sc readelf} &
        \centering{\hspace{0.6cm}(c) \sc libxml} & {\hspace{0.7cm}(d) \sc tiff2pdf}
    \end{tabular}
\caption{Overlap of code coverage between \afledge/\pfuzz and \afl in the 120-hour tests.}
\label{fig:code-cov-overlap-120h}
\end{figure}

\noindent\textbf{Effectiveness of preserving fuzzing capacity.} As discussed in \S~\ref{sec:app}, \afledge can skip certain seeds produced by \afl. This may alter the fuzzing behaviors and, more concerningly, hurt the fuzzing capacity (\ie missing edges that can be covered by vanilla \afl). To understand the impacts of \afledge to the fuzzing capacity, we perform another analysis where we examine whether \afledge and \afl are exploring different edges. Technically, we measure how many of the edges covered by \afl are also covered by \afledge. We show the results in Table~\ref{tab:afl:stat} (the column of \textit{edge overlap rate}). In summary, \afledge can prevalently cover more than 95\% of the edges that are covered by \afl. Considering the existence of randomness, we believe such results strongly support that \afledge largely preserves the behaviors of the vanilla tools and does not significantly affect the fuzzing capacity. 

In comparison to existing solutions (see Table~\ref{tab:afl:stat}), \afledge can preserve as much edge coverage as \pfuzz. This further proves that \afledge well maintains the fuzzing space since \pfuzz does not skip seeds and thus, its results represent the best efforts. Further, \afledge outperforms \pafl in covering the edges reached by \afl (96.2\% \vs 91.8\%). This is because \afledge keeps seeds to cover all the original code to avoid losing fuzzing capacity while \pafl more aggressively skips seeds. 

To validate the above observations in longer-term fuzzing, we extend the tests of \afledge to 120 hours. We also run this test with \pfuzz as a comparison. As shown in Fig.~\ref{fig:code-cov-overlap-120h}, \afledge consistently preserves the edges covered by \afl across the 120 hours, producing results comparable to \pfuzz. We note that in certain cases, \afledge even slightly outperforms \pfuzz. This is mostly because \afledge has a higher efficiency of edge coverage than \pfuzz and therefore, reaches more edges that \afl covers. 


\begin{table}[t!]
\centering
\caption{Impacts of frequency of our task distribution.}
\label{tab:dis:freq}
\begin{tabular}{wc{2cm}|wc{2cm}||c|c|c|c}
\toprule[0.5pt]
\toprule[0.5pt]
\multirow{2}{*}{\textbf{\normalsize{Prog.}}} & 
\multirow{2}{*}{\textbf{\normalsize{Setting}}} &
\multicolumn{4}{c}{\textbf{\normalsize{Number of Edges Covered in 24h}}}
\\ \cline{3-6} && \textit{once / 1h} & \textit{once / 2h} & \textit{once / 4h} & \textit{dynamic}\\\hline

{\sc objdump } & {\sc afl-edge}               & 33422     & 33828   & 33370 &  \textbf{34402}    \\
\hline
{\sc readelf} & {\sc afl-edge}                  & 50932      & 53036 & 51927 & \textbf{53839}       \\
\hline
\toprule[0.5pt]
\toprule[0.5pt]      
\end{tabular}
\end{table}

\noindent\textbf{Impacts of frequency of task distribution.} Recall that \afledge needs to periodically distribute tasks (\S~\ref{sec:app}). Our hypothesis is that the frequency of distribution can affect the effectiveness of our solution and we dynamically adjust this frequency based on the growth of edges. To validate our hypothesis and demonstrate the utility of our dynamic approach, we perform another experiment where we run one round of distribution per 1 hour, 2 hours, and 4 hours. In Table~\ref{tab:dis:freq}, we present the results. It shows that the frequency of distribution truly makes a difference and our dynamic adjustment indeed outperforms solutions with a fixed frequency.

\begin{table}
\centering
\caption{Comparison of seed distillation algorithms. The numbers show the amount of seeds picked by different algorithms. \label{tab:seedistill}}

 %
\begin{tabular}{c||c|c|c|c}
\toprule[0.5pt]
\toprule[0.5pt]
\multirow{2}{*}{\textbf{{Prog.}}} & \multicolumn{4}{c}{\textbf{\normalsize Number of seeds picked after distillation}} \\\cline{2-5} 

& {\tt\scriptsize Unweighted} & {\tt\scriptsize Weight-Time} & {\tt \scriptsize Weight-Size(cmin)} & {\tt\scriptsize Ours}\\ \hline
{\sc objdump}           &  601        &  803         &  599               &  596      \\ \hline
{\sc readelf}           &  939        &  980         &  938               &  683      \\ \hline
{\sc tcpdump}           &  765        &  849         &  756               &  592      \\ \hline
{\sc xml}               &  689        &  862         &  686               &  434      \\ \hline
{\sc nasm}              &  490        &  724         &  492               &  452      \\ \hline
{\sc nm}                &  541        &  701         &  788               &  778      \\ \hline
{\sc tiff2pdf}          &  785        &  920         &  778               &  778      \\ \hline
{\sc tiff2ps}           &  822        &  916         &  817               &  802      \\ \hline
{\sc ffmpeg}            &  593        &  742         &  589               &  581      \\ 
\toprule[0.5pt]
\toprule[0.5pt]

\end{tabular}

\end{table}

\noindent\textbf{Effectiveness of seed distribution.} In our algorithm of task distribution algorithm (Algorithm~\ref{alg:taskdistribute}), the core idea is to pick a subset of seeds that cover the original edges, commonly known as {\em seed distillation}. Past efforts have developed several other seed distillation algorithms, including AFL-CMIN~\cite{aflaflcm94:online} (notated as \emph{Size-Weighted}) and its variants (including ~\cite{Abdelnur2010Spectral}, notated as \emph{Unweighted}; ~\cite{seedoptimize,woo2013black-box}, notated as \emph{Time-Weighted}). Details of the algorithms are as follows.
\begin{itemize}
    \item \noindent\textbf{Unweighted Algorithm} This algorithm always picks a seed whose edges overlap with the non-covered edges the most. It repeats until all edges are covered. 

 \item \noindent\textbf{Time-Weighted} This algorithm iterates each non-covered edge and picks the seed with the shortest execution time to cover the edge, repeating this process until all edges are covered. 

 \item \noindent\textbf{Size-Weighted (AFL-CMIN)} This algorithm iterates each non-covered edge and picks the seed with the smallest size to cover the edge, repeating this process until all edges are covered.

\end{itemize}

We conduct an experiment to compare our algorithm with the existing algorithms: we run these algorithms on 1,000 random seeds from each of our benchmark programs and count the number of picked seeds. As shown in Table~\ref{tab:seedistill}, our algorithm reduces more seeds than all the existing algorithms in every benchmark program, demonstrating better effectiveness. Note that we skipped some other algorithms (\eg ~\cite{hayes2019moon}) as they cannot ensure all original edges (or hit counts of edges) are preserved.

\subsection{Evaluation of Bug Finding}
\label{subsec:eval:bugs}

In the course of evaluation, the fuzzing tools also trigger many crashes. We triage these crashes with AddressSanitizer~\cite{asan} and then perform a manual analysis to understand the root causes. As shown in Table~\ref{tab:crash-table}, \afledge triggers 6,792 unique crashes and 14 previously unknown bugs, outperforming both \pfuzz and \pafl. Moreover, all the bugs detected by \afl and \pfuzz are also detected by \afledge. 

 We also extended the evaluation of bug finding with the LAVA vulnerability benchmark~\cite{dolan2016lava}. However, we omitted the reporting of the results. Basically, our tests with \afl only trigger 1 LAVA bug, regardless of the parallel fuzzing solutions. The major reason is that all LAVA bugs require a four-byte unit in the input to match a random integer value, which is hard to be satisfied by \afl's mutations. 

\begin{table}[t!]
\centering
\caption{Unique crashes / bugs discovered in our tests.}
\label{tab:crash-table}
\begin{tabular}{wc{1.8cm}||wc{1cm}|wc{0.5cm}|wc{1.3cm}|wc{0.5cm}|wc{1.3cm}|wc{0.5cm}
|wc{1.3cm}|wc{0.5cm}|wc{2.2cm}}
\toprule[0.5pt]
\toprule[0.5pt]
\multirow{2}{*}{\textbf{\normalsize{Prog.}}} & \multicolumn{2}{wc{1cm}|}{\textbf{AFL}}      & \multicolumn{2}{wc{1cm}|}{\textbf{PAFL}}     & \multicolumn{2}{{wc{1cm}|}}{\textbf{P-FUZZ}}  &\multicolumn{2}{wc{1.2cm}|}{\textbf{AFL-EDGE}} & \multicolumn{1}{c}{\textbf{Bug Types}}  \\ \cline{2-10} 
                         & {\tt\scriptsize Crash} & {\tt\scriptsize Bug}  & {\tt\scriptsize Crash} & {\tt\scriptsize Bug}  & {\tt\scriptsize Crash} & {\tt\scriptsize Bug} & {\tt\scriptsize Crash} & {\tt\scriptsize Bug} & {\tt\scriptsize ---} \\ \hline 
{\sc ffmpeg}                   &12       &  1   & 72      &  1       & 54      &  1         &  672 & 1 & \scriptsize{heap overflow}      \\\hline

{\sc tiff2pdf}                  & 0      &  0                 &10       & 1           &2       & 1          &  99 & 1  &\scriptsize{failed allocation}       \\\hline
{\sc tiff2ps}                 &  0     &  0                & 0      &  0      &  126     &  1         &  260 & 3  &\scriptsize{heap overflow}      \\\hline

{\sc nasm}                     & 631      & 2                & 1872 & 6                & 1,430       & 6         &5,765        & 9    & \makecell{ \scriptsize{memory leaks} \\
\scriptsize{stack overflow}}     \\ \hline
\textbf{Total}                    &\textbf{643}       &  \textbf{3}      &  \textbf{1954} &  \textbf{8}              & \textbf{1612}      &  \textbf{9}         &\textbf{6792}       & \textbf{14} &  ---  \\
\toprule[0.5pt]
\toprule[0.5pt]
\end{tabular}
\vspace{-1em}
\end{table}

\section{Related Works}
\label{sec:related}



\subsection{Improvements to Algorithms of Grey-box Fuzzing.} Past research has brought three categories of algorithmic improvements to grey-box fuzzing. The first category explores new kinds of {\em feedback} to facilitate seed scheduling and mutation. AFL~\cite{afltech}
considers code branches covered in a round of execution as feedback, which is further refined by Steelix~\cite{li2017steelix}, CollAFL~\cite{collafl}, and PTrix~\cite{ptrix} with more fine-grained, control-flow related information. TaintScope~\cite{taintscope},  Vuzzer~\cite{vuzzer}, GREYONE~\cite{GREYONE}, REDQUEEN~\cite{aschermann2019redqueen}, and Angora~\cite{angora} use taint analysis to identify data flows that can affect code coverage.

The second category of research investigates how to use the above types of feedback to improve code coverage. FairFuzz~\cite{lemieux2018fairfuzz}, GREYONE~\cite{GREYONE}, and ProFuzzer~\cite{profuzzer} rely on the feedback to mutate the existing inputs and derive new ones that have a higher probability of reaching new code. AFLFast~\cite{aflfast}, DigFuzz~\cite{zhao2019send}, and MOPT~\cite{lyu2019mopt} consider the feedback as guidance to schedule inputs for mutation and prioritizes those with higher potentials of leading to new code. 

The last category aims at improving {\em mutations} to remove common barriers that prevent fuzzers from reaching more code. Majundar \etal~\cite{majumdar2007hybrid} introduce the idea of hybrid fuzzing, which
runs concolic execution to solve complex conditions that are
difficult for pure fuzzing to satisfy. The idea was followed and improved 
by many other works~\cite{pak2012hybrid,stephens2016driller,zhao2019send,qsyminsu}. TFuzz transforms target programs
to bypass complex conditions and forces the execution to reach new code. It then uses a validator to reproduce the inputs that meet the conditions in the original program. Angora~\cite{angora} assumes a black-box function at each condition and uses gradient descent to find satisfying inputs, which is later improved by NEUZZ~\cite{neuzz}.


Differing from the above works, our research aims to improve the efficiency of the parallel mode of fuzzing, an orthogonal strategy to facilitate the efficiency of code coverage.

\subsection{Improvements to Execution Speed of Fuzzing.} Beyond algorithmic improvements, other research aims to improve the efficiency of fuzzing by accelerating the fuzzing execution. PTrix~\cite{ptrix}, Honggfuzz~\cite{honggfuzz}, and kAFL~\cite{kafl} use Intel PT~\cite{intel-pt} to efficiently collect control flow data from the
target program. UnTracer~\cite{nagy2019full}, instead of tracing every round of execution, instruments the target programs such that the tracing only starts when new code is reached. RetroWrite~\cite{dinesh2019retrowrite} proposes static binary rewriting to trace code coverage in binary code without heavy dynamic instrumentation. 

\subsection{Improvements to Parallel Fuzzing.} There are two lines of efforts towards better parallel fuzzing in the literature. Xu~\etal~\cite{xu2017designing} design new primitives to mitigate the contention in the file system and extend the scalability of the \textit{fork} system call. These new primitives speed up the execution of the target programs when many instances are running in parallel. This line of efforts facilitates parallel fuzzing from a system perspective, which is orthogonal to our approach. Following the other line, \pfuzz~\cite{song2019p}, \pafl~\cite{liang2018pafl} and Ye~\etal~\cite{ye2019program} propose to distribute fuzzing tasks to different instances to avoid overlaps. We omit the details of \pfuzz and \pafl since they have been discussed in \S~\ref{sec:app} and evaluated in \S~\ref{sec:eval}. The idea of~\cite{ye2019program} is to assign seeds that cover less-visited branches to different instances and further confine the mutations to focus on those branches. In comparison to \afledge, such an idea may skip the exploration of certain code regions and hurt the related fuzzing space. Further, this idea is essentially a variant of \pafl, and thus, we do not compare it with \afledge in our evaluation.

\section{Discussion}
\label{sec:discuss}

In this section, we discuss some of the limitations in our work and  the potential future directions. 

\subsection{Threats to Validity.} The validity of our research faces three threats. First, our research is motivated by the intuition that overlapped mutations can reduce the efficiency of code coverage. Whether such an intuition is correct or not threatens the foundation of our research. To mitigate this threat, as presented in \S~\ref{sec:overview}, we provide empirical evidence to support the fidelity of our intuition through empirical experiments with real-world programs. Second, \afledge skips seeds during task distribution, which by theory may reduce the fuzzing space. To validate this threat, we perform extensive experiments to show that \afledge largely preserves the edge coverage of \afl and thus, avoids hurting the fuzzing space (see \S~\ref{subsec:eval:result}). Finally, \afledge and \afl may detect different bugs and \afledge may miss the bugs detected by \afl. While we provide no theoretical proofs, our empirical evaluation with both real-world programs and standard benchmarks, as shown in \S~\ref{subsec:eval:bugs}, argues against such a threat.



\subsection{More fine-grained Task Distributions are Needed.} \afledge considers a round of mutations to a seed as an individual task. This represents a coarse-grained definition of fuzzing tasks, which can still result in over-laps. For example, we cannot avoid overlapped mutations by different instances to different seeds. For further improvements, an example idea is to adopt more fine-grained definitions of tasks (\eg defining fuzzing tasks based on mutations ~\cite{ye2019program}).

\subsection{Workloads Need to be Considered.} \afledge does not explicitly consider the workloads of different tasks. Instead, it relies on random distribution, expecting to achieve probabilistic equivalent workload assignment. This strategy can be further improved by estimating the workload attached to a seed. For instance, we can do such an estimation based on the size of the seed and the execution complexity of the seed (following the idea of \aflcmin~\cite{aflaflcm94:online} and \qsym~\cite{qsyminsu}). We may also customize the estimation based on how the fuzzing tools determine the mutation cycles.

\section{Conclusion}
\label{sec:conclusion}

This paper focuses on the problem of parallel fuzzing. It presents a study to understand the limitations of the parallel mode in the existing grey-box fuzzing tools. Motivated by the study, we propose a general model to describe parallel fuzzing. This model distributes mutually-exclusive yet similarly-weighted tasks to different instances, facilitating concurrency and also fairness across instances. Guided by our model, we present a novel solution to improve the parallel mode in \afl. During fuzzing, our solution periodically distributes seeds that carry non-overlapped and similarly-weighted tasks to different instances, maximally meeting the requirements of our model. We have implemented our solution on top of \afl and we have evaluated our implementation with \afl on 9 widely used benchmark programs. Our evaluation shows that our solution can significantly reduce the overlaps and hence, accelerate the code coverage.

  \section*{Acknowledgments}

We  would  like  to  thank  the anonymous reviewers for their feedback. This project was supported by NSF  (Grant \#:  CNS-2031377).  Any  opinions,  findings,  and conclusions  or  recommendations  expressed  in  this  paper  are those  of  the  authors  and  do  not  necessarily  reflect  the  views of the funding agency.

\newpage
\renewcommand{\bibsection}{\large \noindent{\textbf{References}}}
\def\bibfont{\scriptsize}
\bibliographystyle{splncs04}
 {\small \bibliography{ref}}

\end{document}